\pdfoutput=1
\documentclass[a4paper,fleqn,usenatbib]{mnras}
\usepackage{amsmath}
\usepackage{tablefootnote}
\usepackage{natbib}
\usepackage{bm}
\usepackage{verbatim}
\usepackage{graphicx}
\usepackage{mathtools, cuted}
\usepackage[T1]{fontenc}
\usepackage{lmodern}
\usepackage{amssymb}
\usepackage{flushend}
\makeatletter
\patchcmd{\NAT@citex}
  {\@citea\NAT@hyper@{%
     \NAT@nmfmt{\NAT@nm}%
     \hyper@natlinkbreak{\NAT@aysep\NAT@spacechar}{\@citeb\@extra@b@citeb}%
     \NAT@date}}
  {\@citea\NAT@nmfmt{\NAT@nm}%
   \NAT@aysep\NAT@spacechar\NAT@hyper@{\NAT@date}}{}{}

\patchcmd{\NAT@citex}
  {\@citea\NAT@hyper@{%
     \NAT@nmfmt{\NAT@nm}%
     \hyper@natlinkbreak{\NAT@spacechar\NAT@@open\if*#1*\else#1\NAT@spacechar\fi}%
       {\@citeb\@extra@b@citeb}%
     \NAT@date}}
  {\@citea\NAT@nmfmt{\NAT@nm}%
   \NAT@spacechar\NAT@@open\if*#1*\else#1\NAT@spacechar\fi\NAT@hyper@{\NAT@date}}
  {}{}

\makeatother

\defcitealias{caldwell2017dust}{CHK17}
\defcitealias{goldreich1995toward}{GS95}
\defcitealias{kandel2017can}{KLP17}

\title[Statistical properties of galactic foregrounds]{Statistical properties of galactic CMB foregrounds: dust and synchrotron
}

\author[Kandel, Lazarian \& Pogosyan]{
D. Kandel,$^{1,2}$
A. Lazarian$^{3}$
and D. Pogosyan$^{2}$
\\
$^{1}$Kavli Institute for Particle Astrophysics and Cosmology, Stanford University, Stanford, CA 94309, USA\\
$^{2}$Physics Department, University of Alberta, Edmonton, T6G 2E1, Canada\\
$^{3}$Department of Astronomy, University of Wisconsin, 475 North Charter Street, Madison, WI 53706, USA\\
}

\date{\today}

\pubyear{2017}

\begin{document}
\label{firstpage}
\pagerange{\pageref{firstpage}--\pageref{lastpage}}
\maketitle

\begin{abstract}
Recent {\it Planck} observations have revealed some of the important statistical properties of synchrotron and dust polarisation, namely, the $B$ to $E$ mode power and temperature-$E$ (TE) mode cross-correlation. In this paper, we extend our analysis in \citet{kandel2017can} that studied $B$ to $E$ mode power ratio for polarised dust emission to include TE cross-correlation and develop an analogous formalism for synchrotron signal, all using a realistic model of magnetohydrodynamical (MHD) turbulence. Our results suggest that the {\it Planck} results for both synchrotron and dust polarisation can be understood if the turbulence in the Galaxy is sufficiently sub-Alfv\'enic. We also show how $B$ to $E$ ratio as well as the TE cross-correlation can be used to study media magnetisation, compressibility, and level of density-magnetic field correlation.
\end{abstract}
\section{Introduction}
The dominant CMB foreground at frequencies $\lesssim 100$ GHz comes from synchrotron emission, which is produced by the interaction of cosmic rays with the Galactic magnetic field lines, and its polarisation is orthogonal both to the line-of-sight (LOS) and to the magnetic field direction (Ribicki \& Lightman 1979). On the other hand, at frequencies $\gtrsim 100$ GHz, this dominant cosmic microwave background (CMB) foreground is thermal emission from oblate dust grains, which are aligned with their longest axes perpendicular to the magnetic field. As the dust emission is more efficient along its longest axes, a linear polarisation orthogonal to magnetic field direction and the LOS is generated. The fluctuation of polarised synchrotron emission, and therefore its polarisation properties, depends on magnetic-field fluctuation, whereas the fluctuation of polarised dust emission depends on both the fluctuation of magnetic-field and dust density. As the ISM is turbulent (see \citealt{armstrong1995electron}; \citealt{chepurnov2010extending}; see also \citealt{2007ARA&A..45..565M} for a review), these fluctuations naturally arise due to the turbulent motions in the ISM. 

Recent {\it Planck} observations have revealed several important statistical properties of the dust \citep{2016A&A...586A.133P} and synchrotron \citep{adam2016planck} polarisation in terms of the angular power spectrum at the intermediate and high Galactic latitudes. This paper attempts to connect the {\it Planck} measurements with theoretical predictions, using magnetohydrodynamic (MHD) model of turbulence.

To study polarisation properties, it is often convenient to decompose a polarisation map into two rotational invariant modes: E, B, and T (temperature) modes. For a randomly oriented map, E and B modes are expected to have equal power. However, observations of dust polarisation in the intermediate to high latitude of our Galaxy suggest that the power in B mode is half of that in E mode \citep{2016A&A...586A.133P}. Similarly, observations of synchrotron polarisation suggest that the power in B mode is about one-third of that in E mode \citep{adam2016planck}. \cite[hereafter \citetalias{caldwell2017dust}]{caldwell2017dust} and \citet[hereafter \citetalias{kandel2017can}]{kandel2017can} have attempted to explain the results {\it Planck} observation of dust polarisation in the context of MHD turbulence. In particular, \citetalias{kandel2017can} concluded that the observed E/B power ratio of dust emission is consistent with the MHD turbulence model as long as turbulence in the intermediate to high Galactic latitudes is sub-Alfv\'enic. In addition, the recent paper \citep{akrami2018planck} suggest detection of positive value  $\sim 0.36$  for the TE cross-correlation in Planck dust emission maps. That gives another handle for contrasting the models to observations.

In this paper, we extend our previous study in \citetalias{kandel2017can} so as to gain a complete picture of statistics of dust polarisation, and develop a similar statistical framework to study synchrotron polarisation in our Galaxy. Relative to 
\citetalias{kandel2017can} our new contributions in this paper
are the study of temperature and E-polarisation (TE) correlation for dust emission and
complete study of B/E power ratio and TE correlations for the  Galactic synchrotron. As we will show later in this paper, TE cross-correlation for dust and synchrotron polarisation can be a unique probe to studying media compressibility and MHD mode contributions, and testing the density-magnetic field correlation in the ISM. Moreover, we show how {\it Planck}'s observed results for B/E powers of dust and synchrotron polarisation are consistent with the model of MHD turbulence.

The study of dust and synchrotron polarised emission is of particular interest in the context of CMB foregrounds (see \citealt{2011ApJS..192...15G}). At angular scales between $10'$ and a few tens of degrees, cosmological B-mode polarisation signals imprinted during the epoch of inflation may be present. As gravitational waves in the inflationary era of our universe produce these modes, the detection of a primordial B-mode polarisation signature would be a significant discovery, and is one of the major scientific goals of many CMB experiments like BICEP2 \citep{2014PhRvL.112x1101B}, Keck-array \citep{2012JLTP..167..827S}, POLARBEAR \citep{arnold2010polarbear}. Since the E and B mode signals from inflation are expected to be small in amplitude, accurate measurement and subtraction of foreground contamination is important to the measurement of primordial E and B mode polarisation. Therefore, theoretical modelling such as what will be developed in this paper will be useful in understanding and properly subtracting CMB foregrounds.

The structure of the paper is the following. In Sec. \ref{mhdmodes}, we describe MHD turbulence as a mix of three MHD modes and formulate the density and magnetic field fluctuations induced by these modes. We present the general formalism for E/B decomposition of a polarisation map in Sec. \ref{sec:gener}. In Sec. \ref{dpluss}, we present a model we adopt in this paper for E and B polarisation amplitudes, as well as temperature fluctuation amplitude induced by fluctuations of dust and synchrotron emission. In Sec. \ref{results}, we present our results for B/E ratios and TE cross-correlations of synchrotron and dust polarisation. Finally, we discuss the  possible implications of our results in Sec. \ref{discuss}.

\section{Mode description of MHD turbulence}\label{mhdmodes}
In this section we briefly explain the physical foundations of the MHD model and define the quantities that are used for our calculation in the paper. 

The work of \citet[hereafter \citetalias{goldreich1995toward}]{goldreich1995toward} has pushed substantial progress in the theory of incompressible MHD turbulence. The \citetalias{goldreich1995toward} model predicts a Kolmogorov velocity spectrum and scale-dependent anisotropy, and these predictions have been confirmed numerically (\citealt{cho2000anisotropy}; \citealt{maron2001simulations}), and are in good agreement with observations (see \citealt{cho2003mhd}). An important property of incompressible MHD turbulence is  that fluid motions perpendicular to the magnetic field are identical to hydrodynamic motions, thus providing a physical insight as to why in some respect MHD turbulence and hydrodynamic turbulence are similar, while in other respect they are different.

MHD turbulence is in general compressible. The description of incompressible MHD turbulence was extended to account for compressibility of turbulent media in \citet{cho2002compressible, cho2003compressible} by decomposing motions into basic MHD modes (Alfv\'en, slow and fast). The Alfv\'enic and slow modes keep the scaling and anisotropy of the incompressible MHD, while fast modes shows different scaling and exhibits isotropy in power spectrum.

The properties of MHD turbulence depend on the degree of magnetization, which can be characterised by the Alfv\'en Mach number $M_A = V_L/a$, where $V_L$ is the injection velocity at the scale $L$ and $a$ is the Alfv\'en velocity. For super-Alfv\'enic turbulence, i.e. $M_A\gg 1$, magnetic forces should not be important in the vicinity of injection scale, thus at these scales, turbulence statistics are effectively isotropic. For sub-Alfv\'enic turbulence, i.e. $M_A<1$, magnetic forces are important, and turbulence statistics is highly anisotropic. Another important parameter is the plasma $\beta$ ($\equiv P_{\text{gas}}/P_{\text{mag}}$), which characterises compressibility of a gas cloud. Formally, $\beta\rightarrow\infty$ denotes incompressible regime.

Now we briefly describe statistical properties of the fluctuations of magnetic and density fields induced by motions in three MHD modes, and define the quantities that are used in our calculations in this paper. The line of sight (LOS) is assumed to be along the $z$ axis, and the mean field $\bm{H}_0=H_0(\sin\theta, 0, \cos\theta)$ is assumed to be aligned in the $x-z$ plane making an angle $\theta$ with the LOS. We consider perturbations with two-dimensional wavevector $\bm{K}=K(\cos\psi, \sin\psi, 0)$ in the $x-y$ plane of the sky, as observations effectively give the two dimensional sky maps. The angle $\alpha$ between wavevector and magnetic field is then $\cos\alpha=\sin\theta\cos\psi$. 

The power-spectrum of magnetic field perturbation for Alfv\'en, slow and fast modes are given by (see \citealt{cho2002compressible}; \citetalias{caldwell2017dust})
\begin{equation}\label{pahk}
P_{a,H}(k, \alpha)=\frac{1}{a^2}F_a(\alpha)P_a(k)~,
\end{equation}
\begin{equation}\label{pshk}
P_{s,H}(k, \alpha)=\frac{2}{a^2D_{-+}}\frac{ F_s(\alpha)P_s(k)}{(1+D^2_{++}/D^2_{--}\tan^2\alpha)}~,
\end{equation}
and
\begin{equation}\label{pfhk}
P_{f,H}(k, \alpha)=\frac{2}{a^2D_{++}}\frac{ F_f(\alpha)P_f(k)}{(1+D_{-+}^2/D_{+-}^2\tan^2\alpha)}~,
\end{equation}
where $a\equiv H_0/\sqrt{4\pi \rho_0}$ is the Alfv\'en speed and $D_{\pm\pm}=1\pm\sqrt{D}\pm\beta/2~,$ and $D=(1+\beta/2)^2-2\beta\cos^2\alpha~.$

In Eqs. \eqref{pahk}, \eqref{pshk} and \eqref{pfhk}, $P(k)F(\alpha)$ describes anisotropic power, which contains factorised scale dependent part and an angle dependent part. 
The anisotropic dependence of the power spectrum applicable for Alfv\'en and slow modes is given by (\citetalias{goldreich1995toward})
\begin{equation}\label{anisgs}
F_{\text{a,s}}(\alpha)=\exp\left[-\frac{M_A^{-4/3}\lvert\cos\alpha\rvert}{(\sin^2\alpha)^{1/3}}\right]~.
\end{equation}
Note that the model of $F_\text{a,s}(\alpha)$ used in \citetalias{kandel2017can} omit the square root of the angular argument. However, this does not noticeably change the results for dust polarisation parameters. 

Since the power spectrum of fast mode is isotropic, we take
$F_\text{f}(\alpha)=1$ from now on.

In order to study dust polarisation, we also need to define power spectrum of the density fluctuations. Within our model, Alfv\'en modes cannot induce any density fluctuations, the only contribution to density fluctuations comes from fast and slow compressible modes. Under the assumption of frozen-in magnetic field, the power spectrum of fractional density fluctuations for slow and fast modes can be written as (see \citetalias{kandel2017can})

\begin{equation}\label{psrk}
P_{s,\rho}(k, \alpha)=\frac{2}{a^2D_{-+}}\frac{\sin^2 2 \alpha \,F_s(\alpha)P_s(k)}{(D^2_{--}\cos^2\alpha+D^2_{++}\sin^2\alpha)}~,
\end{equation}
and
\begin{equation}\label{pfrk}
P_{f,\rho}(k, \alpha)=\frac{2}{a^2D_{++}}\frac{\sin^2 2\alpha\, F_f(\alpha)P_f(k)}{(D^2_{+-}\cos^2\alpha+D^2_{-+}\sin^2\alpha)}~.
\end{equation}

In order to study dust polarisation properties, besides the power spectrum of density field and magnetic field, one also needs knowledge of density-magnetic field correlation. Following our development in \citetalias{kandel2017can}, we assume that  density and magnetic field are uncorrelated. A detailed justification of the validity of this approximation is
based on the process of fast magnetic reconnection in turbulent fluids \citet{lazarian1999reconnection} and has been supported by the observations of the marginal correlation of density and magnetic field strength in diffuse media (see \citealt{crutcher2010magnetic}). We provide more justification of this point in \citetalias{kandel2017can}.  This is different from \citetalias{caldwell2017dust} who have assumed the perfect flux-freezing condition.
In \citetalias{kandel2017can} we have shown that the difference between
two models is small for $E/B$ power ratio. However later in this paper,
we'll show that the $\langle TE \rangle$ correlation has sensitivity and discriminatory power to the correlation between the dust density and the magnetic field.

In summary, the model of MHD turbulence we presented has the following main parameters: $M_A$, and $\beta$. Anisotropy is highly sensitive to $M_A$ (see equation \ref{anisgs}).  Ultimately, observational data should be used to place limits in space of physical parameters, $M_A$ and $\beta$. Another consideration that needs to be made is the model of density fluctuations, and their correlation with the magnetic field.

\section{General formalism of $E$ and $B$ decomposition of polarisation map and their powers}\label{sec:gener}
In this section, we briefly give details of the mathematical characterisation of linear polarisation maps, and various polarisation parameters that are used to characterise these maps. 

The linearly polarised emission is conveniently described by Stokes parameters $I(\bm{\theta})$, $Q(\bm{\theta})$ and $U(\bm{\theta})$, measured with respect to a reference axes in the two-dimensional sky. 
In cosmological context, intensity $I(\bm{\theta})$ is interpreted as
temperature $T(\bm{\theta})$.  polarisation part can also be written as a complex polarisation $\Pi(\bm{\theta})=Q(\bm{\theta})+iU(\bm{\theta})$. 
The complex polarisation, for optically thin emission, is related to the polarised emissivity $\epsilon_{\Pi}$ as

\begin{equation}\label{Pi}
\Pi(\hat{\bm{n}})=\int_0^{\infty}\mathrm{d}r\,\epsilon_{\Pi}(r\hat{\bm{n}})~,
\end{equation}
where $\hat{\bm{n}}$ is the 3-dimensional unit vector pointing towards a given direction in the sky.  
The polarised emissivity can be represented in Fourier space as
\begin{equation}\label{emm}
\epsilon_\Pi(r\hat{\bm{n}})=\int\frac{\mathrm{d}^3k}{(2\pi)^3}\epsilon_{\Pi}(\bm{k})\mathrm{e}^{i\bm{k}\cdot\hat{\bm{n}}r}~,
\end{equation}
where $\bm{k}$ is the three-dimensional wave-vector. 

The Stokes parameters $Q$ and $U$ transform under a right handed rotation $\hat{n}$ by an angle $\alpha$ as
\begin{align}
Q'&=Q\cos 2\alpha + U \sin 2\alpha\nonumber\\
U'&=-Q\sin 2\alpha+U\cos 2\alpha~,
\end{align}
and thus the complex polarisation transform under rotation by an angle $\alpha$ as a spin-2 field:
\begin{equation}
\Pi(\hat{n})\rightarrow \mathrm{e}^{\mp2i\alpha}\Pi(\hat{n})~.
\end{equation}

It is more convenient to represent the polarisation field in terms of rotationally invariant (scalar) modes: $E$ and $B$ modes. 
In the flat sky approximation, the $E$ and $B$ modes are related to $Q$ and $U$ parameters in Fourier space as (see \citealt{kamionkowski2016quest})
\begin{equation}\label{eq:transform}
(\tilde{E}+i\tilde{B})(\bm{\ell})=(\tilde{Q}+i\tilde{U})\mathrm{e}^{-2i\psi_\ell}~,
\end{equation}
where $\psi_\ell$ is the angle $\bm{\ell}$ makes with $\hat{\bm{\theta}}_x$.
The $E-$polarisation is a completely curl-free mode, whereas the $B-$polarisation is completely divergence-free. This decomposition of polarisation map into $E$ and $B$ is particularly useful in cosmological context, as $B$ modes in the early universe can only be produced by gravitational waves from inflation, and thus detection of primordial $B$ modes can be a unique support to the inflation model. 

For a rotationally invariant scalar quantity $\mathcal{Q}=(E,B, T)$, under the assumption of statistical isotropy, one can show that the angular power spectrum of $\mathcal{Q}$, $C_\ell^{\mathcal{Q}\mathcal{Q}}$, is related to the 3D power spectrum of the
emissivity of the same component $P_{\mathcal{Q}\mathcal{Q}}(k)$ as
\begin{align}
C_\ell^{\mathcal{Q}\mathcal{Q}}=\int\mathrm{d}r\,\int\mathrm{d}r'\,\frac{2}{\pi}\int\mathrm{d}k\,k^2 P_{\mathcal{Q}\mathcal{Q}}(k)j_{\ell}(kr)j_{\ell}(kr')~,
\end{align}
where we used the definition of the power spectrum of polarised emissivity of $\mathcal{Q}$
\begin{equation}
\langle\epsilon_{\mathcal{Q}}(\bm{k})\epsilon_{\mathcal{Q}}(\bm{k}')\rangle\equiv (2\pi)^3\delta(\bm{k}-\bm{k}')P_{\mathcal{Q}\mathcal{Q}}(\bm{k})~.
\end{equation}

Assuming $P_{\mathcal{QQ}}$ to be slowly-varying function of $k$, which is the so-called Limber's approximation finally allows us to write 
\begin{equation}\label{maincell}
C_\ell^{\mathcal{Q}\mathcal{Q}}=\int\mathrm{d}r\,\frac{1}{r^2}P_{\mathcal{Q}\mathcal{Q}}\left(k=\frac{l+1/2}{r}\right)~.
\end{equation}

Using equation \eqref{maincell}, one can obtain the power of both $E$ and $B$ polarisation. If one considers a single MHD mode giving rise to fluctuations of polarisation, the ratio of powers in $E$ and $B$ polarisation modes is simply given by the ratio $C^{EE}_{\ell}/C^{BB}_{\ell}$, each averaged suitably over 3D angles, and is independent of $\ell$. However, as it will be shown later, if the one considers fluctuations arising from different MHD modes, this ratio is dependent on $\ell$, albeit weakly.

In addition to power spectra of $T$, $E$, and $B$, polarisation maps are also described by correlations between them. The temperature and $E$ modes are symmetric under reflection, thus have the same parity, whereas the $B$ mode is asymmetric under reflection, thus has odd parity. This immediately means that the power $P_{EB}=P_{TB}=0$, and therefore the temperature/polarisation maps are completely determined by the four powers $P_{EE}, P_{BB}, P_{TT}$ and $P_{TE}$ \citep{kamionkowski2016quest}. 

\section{E and B mode description of dust and synchrotron polarisation}\label{dpluss}
In this section, we present theoretical framework to study dust and synchrotron polarisation in the context of MHD turbulence, which consists of three modes: incompressible Alfv\'en mode, and compressible fast and slow modes. We provide expressions of E and B powers for each MHD mode, for both dust and synchrotron polarisation.
\subsection{E and B mode description of synchrotron and dust polarisation}\label{app:sync}

Synchrotron emission arises due to the motion of relativistic electrons in magnetised regions. The emission is essentially non-linear in the magnetic field $H$, due to relativistic effects. Such emission is linearly polarised.

The observed two-dimensional projection of an emitting volume has polarised emissivity of the following form
\begin{equation}\label{eq:inten}
\epsilon_{\text{sync}}=\epsilon_Q+i\epsilon_U=A_{\text{sync}} H_\perp^{\gamma_{\text{sync}}}(H_x+\mathrm{i}H_y)^2~,
\end{equation}
where $A_{\text{sync}}$ is a normalisation constant that depends on several parameters such as the number density of relativistic electrons, $H_\perp$ is the sky-projected magnetic field which depends on $X = (x, y)$ is the 2D position vector on the sky, and $\gamma_{\text{sync}}$  is, generally, a fractional power. For the galactic synchrotron emission, $\gamma_{\text{sync}}$ is found to take an approximate value of 0 (see \citealt{getmantsev1959determination}; \citealt{chibisov1981angular}; \citealt{lazarian1990correlation}). Since the synchrotron polarisation axis is perpendicular to the sky projected magnetic field \citep{rybicki2008radiative}, we take $A_{\text{sync}}<0$, and as we will see later, this sign is important for TE cross-correlation.

Dust emission, one the other hand, is primarily thermal emission from dust grains with their longest dimension perpendicularly aligned with the magnetic field. Despite being of different origin, dust emission is also linearly polarised with the direction of polarisation reflecting the magnetic field and its emissivity can be cast in a similar form to synchrotron
\begin{equation}\label{pemiss}
\epsilon_P=\epsilon_Q + i\epsilon_U=A_dn_dH^{\gamma_d}(H_x+iH_y)^2~.
\end{equation}
 For $\gamma_d=-2$, the polarised emission is independent of magnetic field strength. In equation \eqref{pemiss}, $n_d$ is dust density which we take to be proportional to gas density (see \citealt{lazarian2002grain}), and $A_d<0$ is a constant, and its value is taken to be negative so that the dust polarisation is perpendicular to the direction of magnetic field.
 
We now write down the expressions for E and B power for each MHD mode for both synchrotron and dust polarisation using Appendix \ref{app:general}. 
Proportionality to inhomogeneous dust density is one of the main differences between
 the dust emission in comparison with synchrotron which
 is proportional to electron density that we consider homogeneous. Similarly, while dust polarisation is dependent on 3d structure of the magnetic field, synchrotron polarisation is sensitive only to the sky projection of magnetic field.

The E and B synchrotron powers are presented in Eqs. \eqref{eq:gaesyncm} - \eqref{gsfmb}. For Alfv\'en modes, we have
\begin{align}\label{eq:maingaesync}
\langle\tilde{E}^2\rangle_{\text{sync},a}
\propto(\sin\theta&)^{2\gamma_{\text{sync}}+2}\cos^2\theta\sin^2\psi\nonumber\\
&\times\left(\frac{2-\gamma_{\text{sync}}\cos2\psi}{\sin\alpha}\right)^2P_{a,H}(\alpha)~,
\end{align}
and
\begin{align}\label{eq:maingabsync}
\langle\tilde{B}^2\rangle_{\text{sync},a}\propto(\sin\theta&)^{2\gamma_{\text{sync}}+2}\cos^2\theta\cos^2\psi\nonumber\\
&\times\left(\frac{2+2\gamma_{\text{sync}}\sin^2\psi}{\sin\alpha}\right)^2P_{a,H}(\alpha)~,
\end{align}
where $P_{a,H}(\alpha)$ is defined in Eq. \eqref{pahk}. Similarly, for slow and fast modes, we have
\begin{align}\label{eq:cesync}
\langle\tilde E^2\rangle_{\text{sync},i}\propto(\sin\theta&)^{2\gamma_{\text{sync}}+4}\sin^4\psi\nonumber\\
&\times\left(\frac{2-\gamma_{\text{sync}}\cos2\psi}{\sin\alpha}\right)^2P_{i,H}(\alpha)~,
\end{align}
and
\begin{align}\label{eq:cbsync}
\langle\tilde B^2\rangle_{\text{sync},i}\propto(\sin\theta&)^{2\gamma_{\text{sync}}+4}\sin^2\psi\cos^2\psi\nonumber\\
&\times\left(\frac{2+2\gamma_{\text{sync}}\sin^2\psi}{\sin\alpha}\right)^2P_{i,H}(\alpha)~,
\end{align}
where $P_{i,H}(\alpha)$ indicates the power spectrum of magnetic field of slow and fast modes, defined in Eqs. \eqref{pshk} and \eqref{pfhk}.
These expressions simplify significantly for $\gamma_{\text{sync}}=0$, which is the observationally motivated index for Galactic synchrotron emission. 

Similarly, for dust polarisation, the E and B powers are given Eqs. \eqref{eq:gaesyncdust} - \eqref{gsfmbdust}. For Alfv\'en modes, we have
\begin{equation}\label{main:gaesyncdust}
\langle\tilde{E}^2\rangle_{d,a}\propto\sin^2 2\theta\frac{\sin^2\psi}{\sin^2\alpha}P_{a,H}(\alpha)~,
\end{equation}
and
\begin{equation}\label{main:gabsyncdust}
\langle\tilde{B}^2\rangle_{d,a}\propto\sin^2 2\theta\frac{\cos^2\psi}{\sin^2\alpha}P_{a,H}(\alpha)~.
\end{equation}
Similarly, for slow and fast modes, with the assumption of uncorrelated density and magnetic field, we have
\begin{align}\label{sfm}
\langle\tilde E^2\rangle_{d,i}\propto  \frac{\sin^4\theta(1-\cos2\psi(1+\gamma_{\text{d}}\sin^2\alpha))^2}{\sin^2\alpha}P_{i, H}(\alpha)\nonumber\\
+\sin^4\theta\cos^22\psi P_{i, \rho}(\alpha)~,
\end{align}
and
\begin{align}\label{sfmb}
\langle\tilde B^2\rangle_{d,i}\propto \frac{\sin^4\theta\sin^22\psi(1+\gamma_{\text{d}}\sin^2\alpha)^2}{\sin^2\alpha}P_{i, H}(\alpha)\nonumber\\
+\sin^4\theta\sin^22\psi P_{i, \rho}(\alpha)~,
\end{align}
where $P_{i,\rho}(\alpha)$ indicates density power spectrum of slow and fast modes, defined in Eqs. \eqref{psrk} and \eqref{pfrk}.

One of the main quantity of interests in this paper is the ratio of B to E power. Since Planck measurement of both dust and synchrotron reveal a rather uniform B/E ratio across different patches of the sky, where magnetic orientations could be rather different, it is meaningful to calculate the angle averaged B/E ratio. Therefore, we follow the procedure of \citetalias{caldwell2017dust}, and define the B/E power ratio as
\begin{equation}\label{eq:powerBE}
R=\frac{\int\mathrm{d}\Omega\langle\tilde{B}^2\rangle}{\int\mathrm{d}\Omega\langle\tilde{E}^2\rangle}~.
\end{equation}
Here $\mathrm{d}\Omega=\sin\theta\,\mathrm{d}\theta\,\mathrm{d}\psi$. Equally, one could carry out the following transformation $\cos\theta=\sin\alpha\cos\varpi, \sin\theta\sin\psi=\sin\alpha\sin\varpi$, and $\sin\theta\cos\psi=\cos\alpha$, such that the differential volume can be equivalently written as $\mathrm{d}\Omega=\sin\alpha\,\mathrm{d}\alpha\,\mathrm{d}\varpi$.

If the ratio is computed for the powers in individual mode, the $k$ dependence 
 In general, MHD turbulence contains a mixture of MHD modes, each with different wave spectrum, and therefore in such cases, one has to use the full expression given by Eq. \eqref{fullequation1}. However, as one can clearly see from Eq. \eqref{fullequation1}, the dependence of the B/E ratio on $\ell$ is very weak, and therefore, we ignore this dependence in our calculations later in this paper, and instead, by assuming equal power in all three MHD modes at injection scale, use 
\begin{equation}\label{fullequation}
R=\frac{P_{a,B}+P_{s,B}+P_{f,B}}{P_{a,E}+P_{s,E}+P_{f,E}}~,
\end{equation}
where $P_{a,E}, P_{a,B}$ are the amplitudes of power of $E$ mode and $B$ mode, respectively, of Alfv\'en mode, and so on for fast and slow modes. 
\subsection{TE cross-correlation for synchrotron and dust polarisation}\label{TE-cross}
Another important measure of polarisation is the temperature-E (TE) correlation. The brightness temperature of the polarised emission is proportional to its intensity. Thus, fluctuations of intensity, which arises from the fluctuations of the magnetic field for synchrotron and from fluctuations of both the magnetic field and the dust density for dust, give rise to temperature fluctuations. Taking intensity of synchrotron polarisation to be $I_{\text{sync}}\propto H_\perp^{\gamma_\text{sync}+2}$, and intensity of dust polarisation to be $I_{\text{dust}}\propto n_d$  we have
\begin{equation}\label{tempsync}
\delta T_{\text{sync}}=c_{\text{sync}}H_0^{\gamma_{\text{sync}}+2}(\sin\theta)^{\gamma_{\text{sync}}+1}\frac{\delta H_\perp}{H_0}~,
\end{equation}
and 
\begin{equation}\label{tempfluct}
\delta T_{\text{dust}}=c_{\text{dust}}\delta n~,
\end{equation}
where $c_{\text{sync}}, c_{\text{dust}}$ are constants of proportionality. Using equation \eqref{tempsync} and \eqref{tempfluct}, one can thus write the Fourier-amplitude of temperature fluctuation projected through a box of width $\Delta r$ as
\begin{equation}
\tilde{T}_{\text{sync}}(\bm{\ell})=cT_0\frac{\Delta r}{r^2}(\sin\theta)^{\gamma_{\text{sync}}+1}\frac{\delta\tilde{H}_x}{H_0}~.
\end{equation}
Re-writing this into two transverse wave-vector modes: $\hat{\bm{a}}$ (Alfv\'en) and $\hat{\bm{\theta}}$ (slow and fast), we have
\begin{equation}\label{tesync}
\tilde{T}_{\text{sync}}(\bm{\ell})\propto (\sin\theta)^{\gamma_{\text{sync}}+1}\frac{\sin\psi}{\sin\alpha}\left(\cos\theta\frac{\delta\tilde{H}_a}{H_0}+\sin\theta\sin\psi\frac{\delta\tilde{H}_p}{H_0}\right)~.
\end{equation}

Similarly
\begin{equation}\label{tempamp}
\tilde{T}_{\text{dust}}(\ell)\propto \frac{\delta\tilde{n}(k)}{n_0}
\end{equation}

From equation \eqref{tesync} and \eqref{tempamp}, one can immediately see that Alfv\'en modes do not contribute to temperature fluctuations for dust polarisation, while they do contribute to synchrotron temperature fluctuations. This apparent asymmetry is due to the fact that while $H$ is solenoidal, its sky projection $H_\perp$ is not. 

Using equations \eqref{esync} and \eqref{tesync}, we get temperature E-mode (TE) and TT correlation for synchrotron due to Alfv\'en mode as
\begin{align}\label{finaltesyncalfven}
\langle TE\rangle_{\text{sync}, a}\propto (\sin\theta)^{2\gamma_{\text{sync}}+2}&\frac{\cos^2\theta\sin^2\psi}{\sin^2\alpha}\nonumber\\
&\times\left(2-\gamma_{\text{sync}}\cos 2\psi\right)P_{a,H}(\alpha)~,
\end{align}
and 
\begin{equation}\label{aTTsync}
\langle TT\rangle_{\text{sync}, a}\propto (\sin\theta)^{2\gamma_{\text{sync}}+2} \frac{\cos^2\theta\sin^2\psi}{\sin^2\alpha}P_{a,H}(\alpha)~.
\end{equation}
For compressible modes, we have
\begin{align}\label{finaltesync}
\langle TE\rangle_{\text{sync}, i}\propto  (\sin\theta)^{2\gamma_{\text{sync}}+4}&\frac{ \sin^4\psi}{\sin^2\alpha}\nonumber\\
&\times\left(2-\gamma_{\text{sync}}\cos 2\psi\right)P_{i,H}(\alpha)~,
\end{align}
and
\begin{equation}\label{cTTsync}
\langle TT\rangle_{\text{sync}, i}\propto (\sin\theta)^{2\gamma_{\text{sync}}+4} \frac{\sin^4\psi}{\sin^2\alpha}P_{i,H}(\alpha)~.
\end{equation}

Similarly, using  equations \eqref{esyncdust} and \eqref{tempamp}, we have the TE correlation for dust polarisation due to compressible MHD modes as
\begin{equation}\label{eq:dte}
\langle TE\rangle_{d,i}\propto \sin^2\theta\cos 2\psi P_{i,\rho}(\alpha)~.
\end{equation}

We are finally interested in the normalised cross-correlation coefficient $r_i$, which is
\begin{equation}\label{tecrossr}
r_i=\frac{\int\mathrm{d}\Omega\, \langle TE\rangle}{\sqrt{\int\mathrm{d}\Omega\, \langle TT\rangle}\sqrt{\int\mathrm{d}\Omega\, \langle EE\rangle}}~.
\end{equation}
The ratio $r$ has been recently measured by {\it Planck}, and will thus be useful in comparing our predictions with data.

\section{Results for various synchrotron and dust polarisation parameters}\label{results}
Here we present most of our important results for synchrotron and dust polarisation parameters: B/E power ratio and TE cross-correlation. The detailed derivation of $E$ and $B$ amplitude of {\it dust polarisation} and E/B power ratio for three MHD modes are presented in our previous letter \citetalias{kandel2017can}, but for completeness, we still present this ratio in this paper, along with E/B power ratio for synchrotron polarisation. 

All the important derivations for E, B mode amplitude, and power as well as temperature amplitude, along with temperature-E mode cross-correlation of dust and synchrotron are presented in the Appendix \ref{dpluss}. Here we present final results.
\subsection{E/B power ratio for synchrotron and dust polarisation}\label{sec:EBpower}
At lower frequency, the dominant foreground for CMB polarisation is the synchrotron polarisation. We numerically characterise the various polarisation parameters of this polarisation for three different MHD modes at different Alfv\'en Mach number, and plasma $\beta$. First, we calculate the amplitude of E, B power and their ratio using Eqs. \eqref{eq:maingaesync}-\eqref{eq:cbsync} with Eqs. \eqref{eq:powerE}, \eqref{eq:powerB} and \eqref{eq:powerBE}. When calculating B/E power ratio of an equal mix of three MHD modes, we assume equal power among each MHD mode at the injection scale. In such case, Eq. \eqref{fullequation1} suggests that the dependence of B/E ratio to $\ell$ is weak, and therefore, we use Eq. \eqref{fullequation}, ignoring the $\ell$ dependence altogether. Angular transformation, which facilitates easier evaluation of angular integrals in these equations, is presented in the final paragraph of Appendix. \ref{app:general}.  

Fig. \ref{figsync} shows the amplitude of E and B power, and their ratio for synchrotron polarisation. The upper and central panels of Fig. \ref{figsync} show that fast mode dominates in amplitude at low $\beta$. This would naturally lead to the dominance of fast mode on the B/E power ratio at the low $\beta$ regime. {\it Planck} measurement of synchrotron emission \citep{adam2016planck} at intermediate to high Galactic latitude suggests a B/E power ratio of $\sim 0.35$.  Since B/E power ratio due to the fast mode is $\sim 0.7$, suppression of fast mode is necessary to explain the observed ratio. This suppression can in part be expected, as most of the synchrotron emission comes from Galactic halo, which is magnetically dominated and incompressible, and thus the "shock-like" pure fast modes are expected to be sub-dominant in such regions of the sky. 
\begin{figure*}
\centering
\includegraphics[scale=0.4]{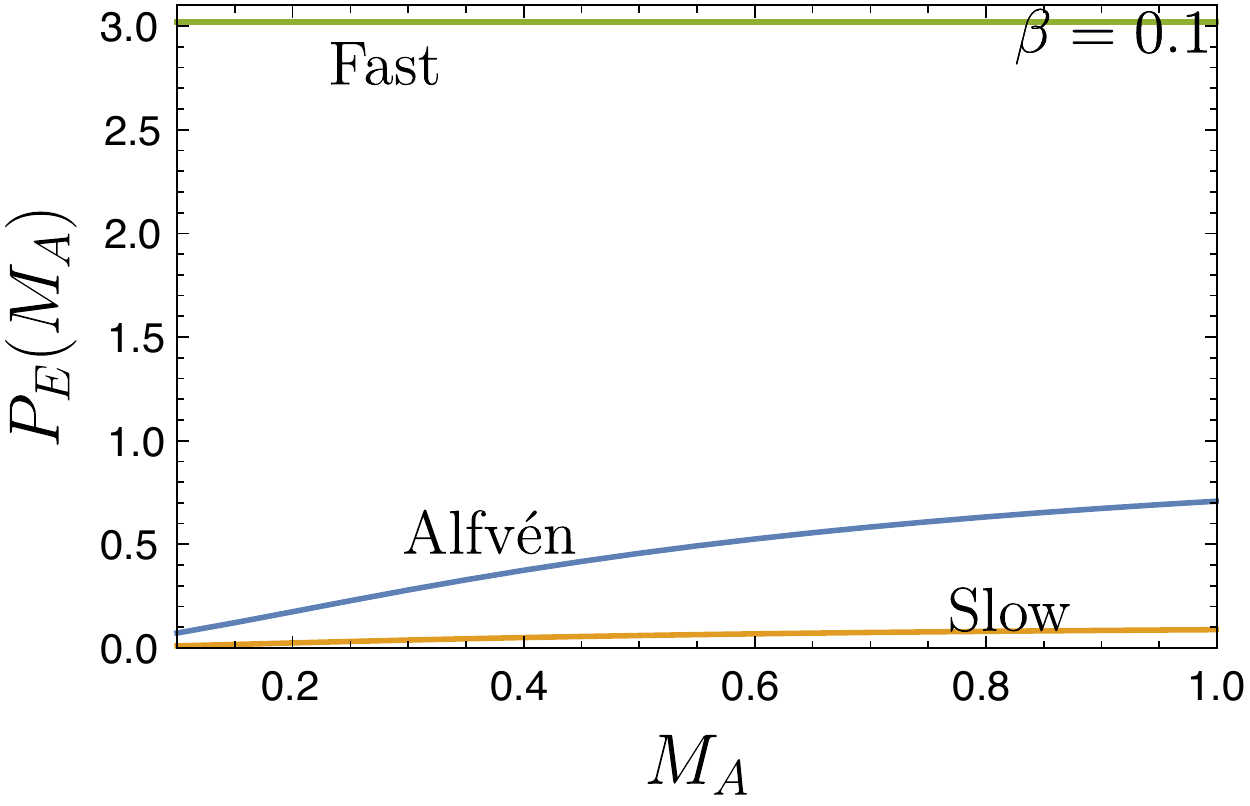}\hspace*{0.1cm}
\includegraphics[scale=0.4]{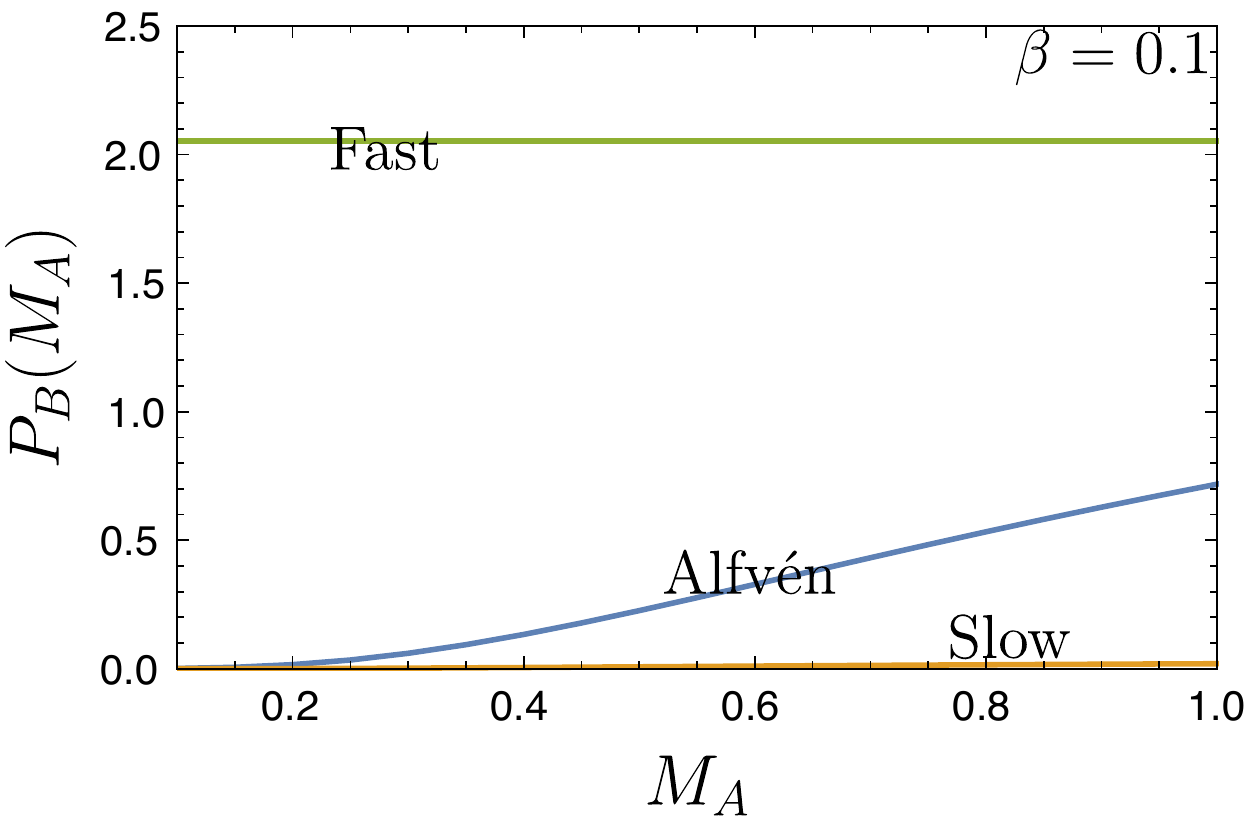}\hspace*{0.1cm}
\includegraphics[scale=0.4]{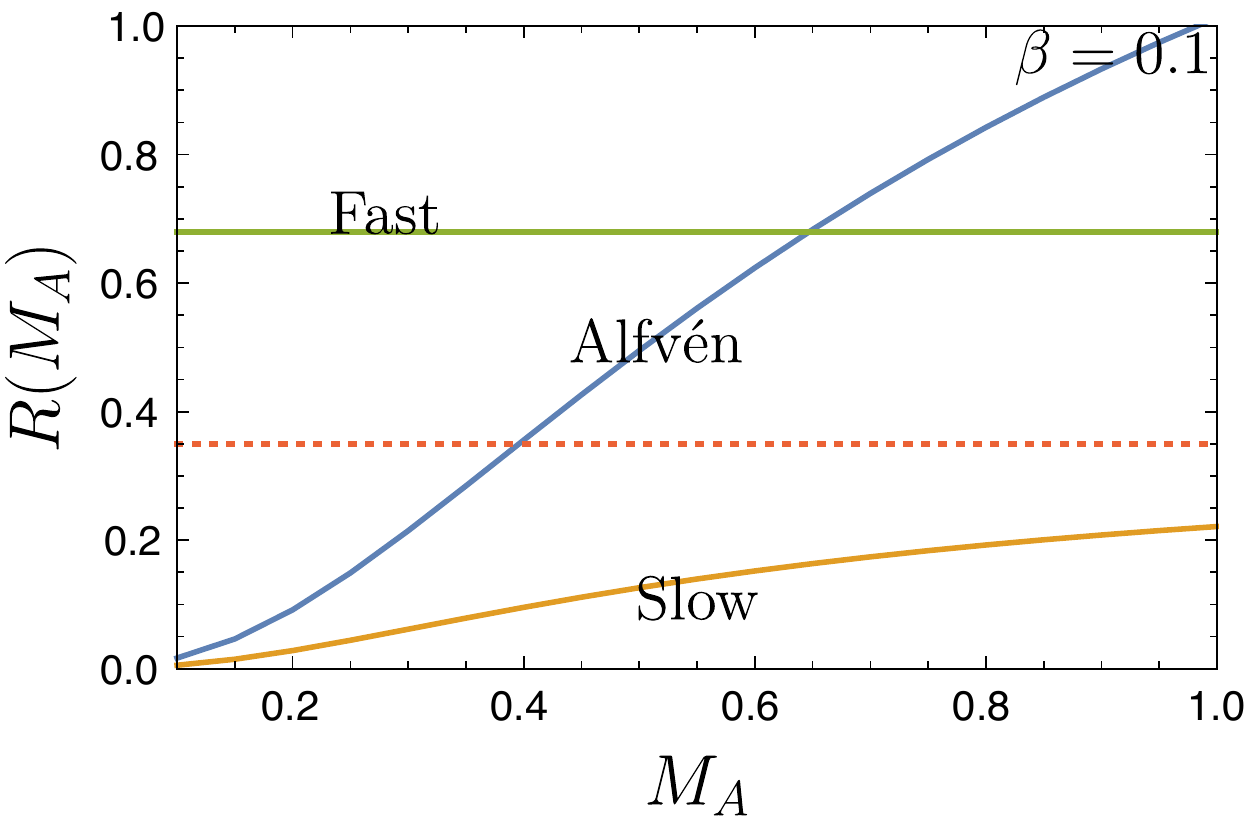}\hspace*{0.1cm}\\
\vspace{0.1cm}
\includegraphics[scale=0.4]{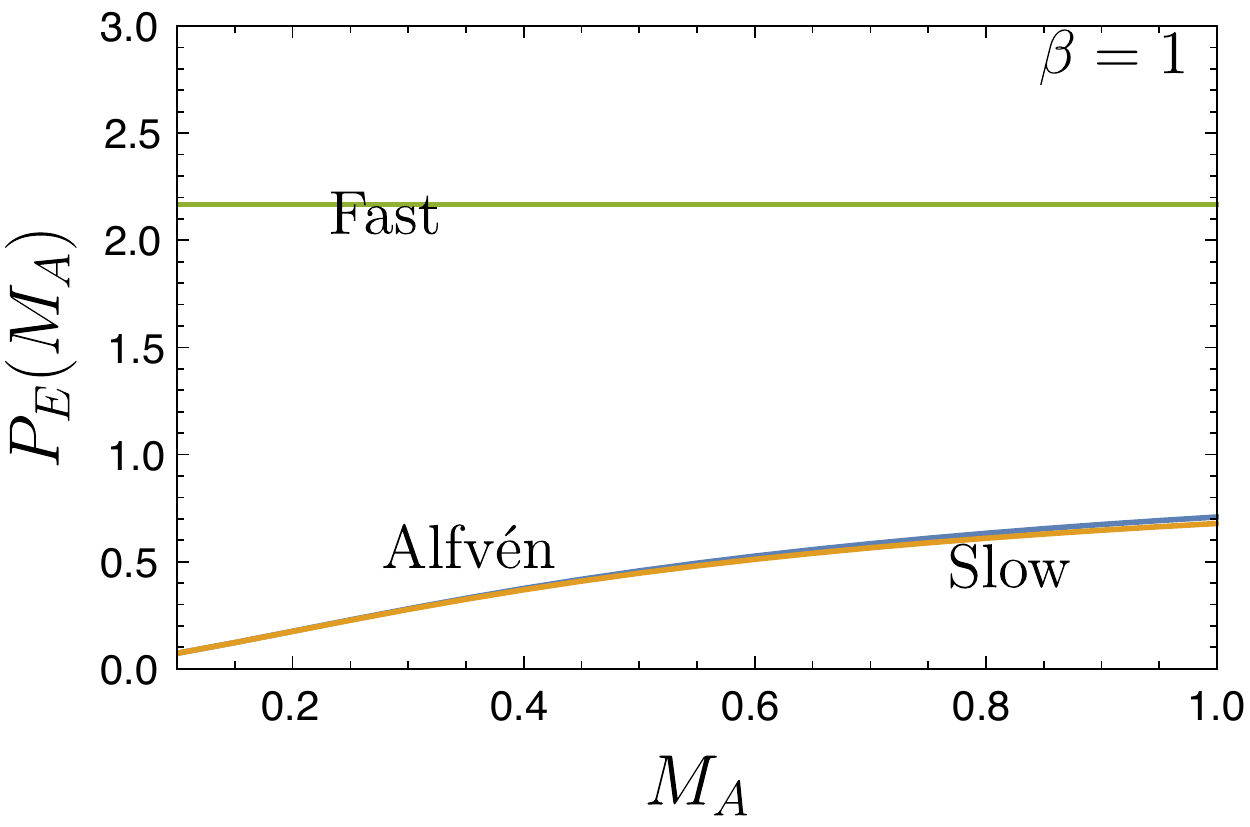}\hspace*{0.1cm}
\includegraphics[scale=0.4]{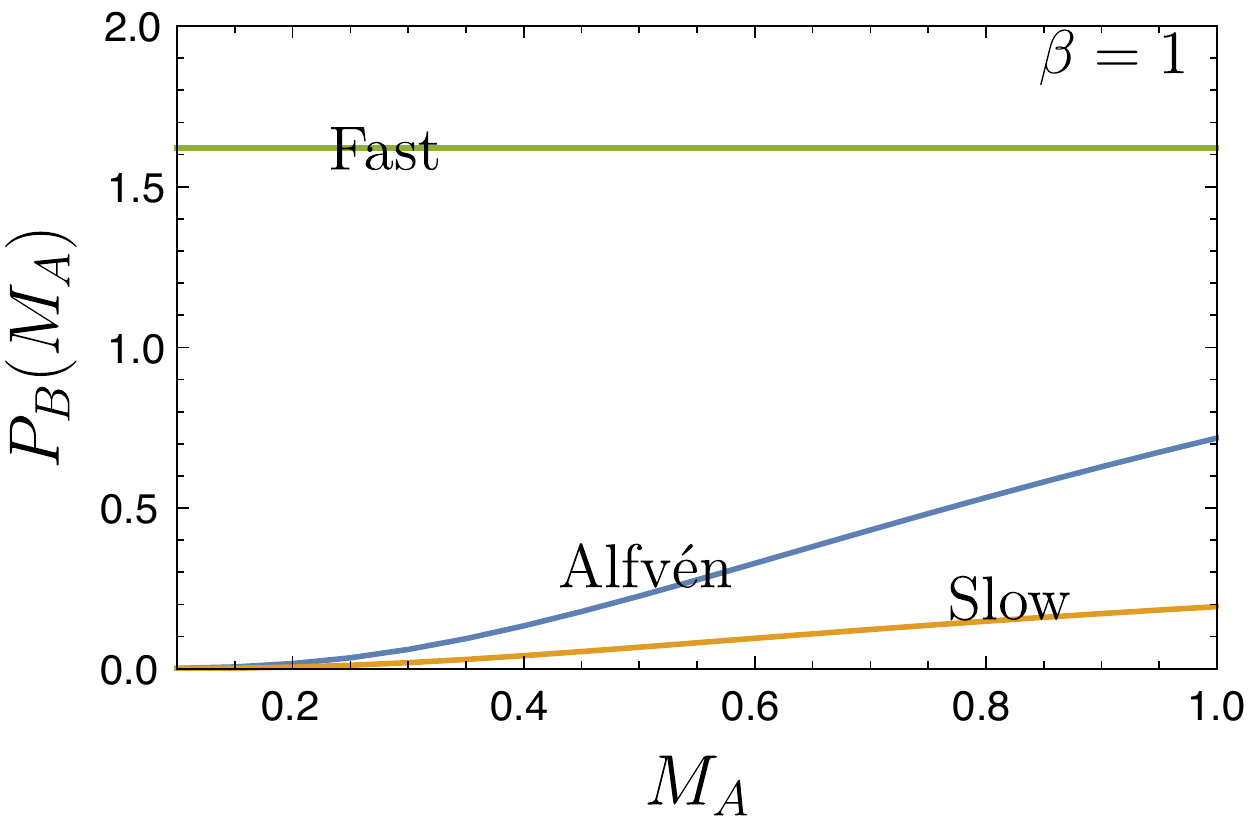}\hspace*{0.1cm}
\includegraphics[scale=0.4]{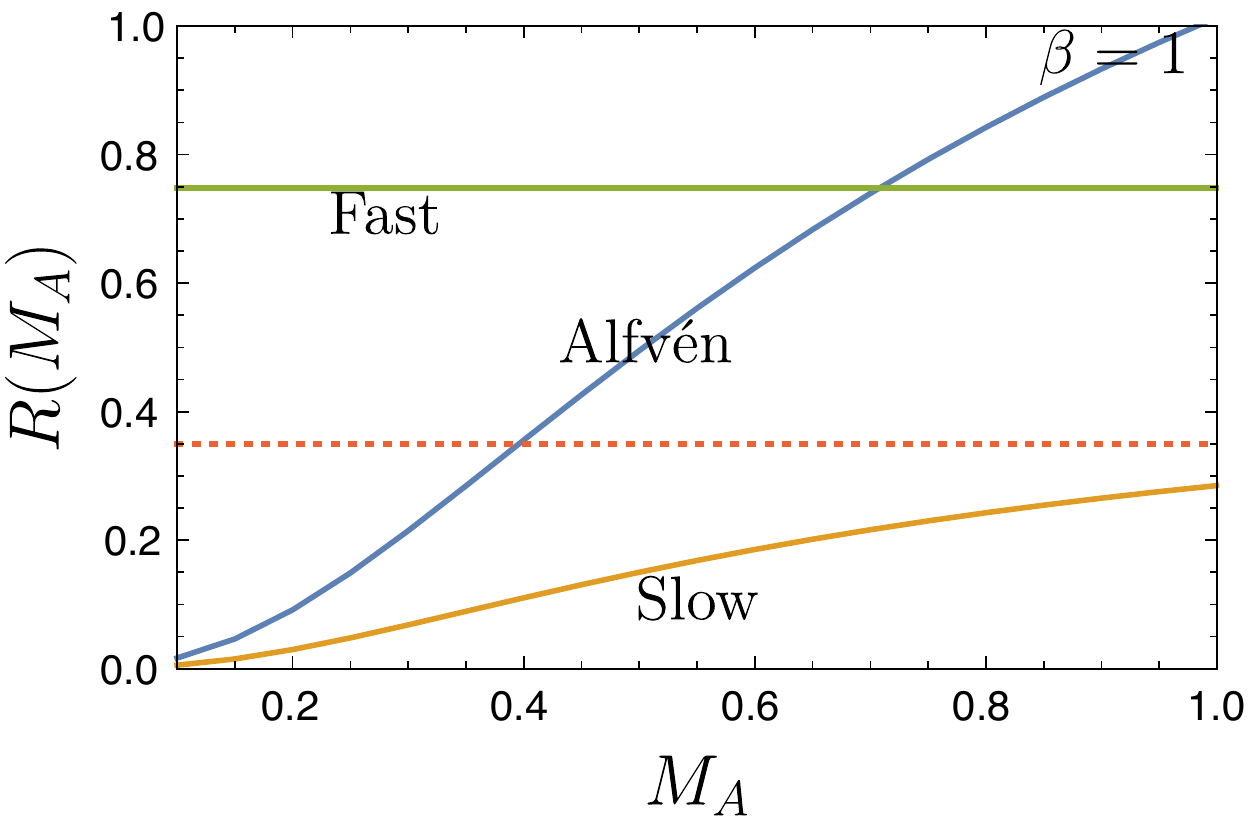}\hspace*{0.1cm}
\caption{$E$ and $B$ amplitudes, and their ratios for synchrotron polarisation, assuming $\gamma_{\text{sync}}=0$. Upper row is for $\beta=0.1$ and lower is for $\beta=1$. The dotted line shows the expected ratio of 0.35.}
\label{figsync}
\end{figure*}

\begin{figure}
\centering
\includegraphics[scale=0.4]{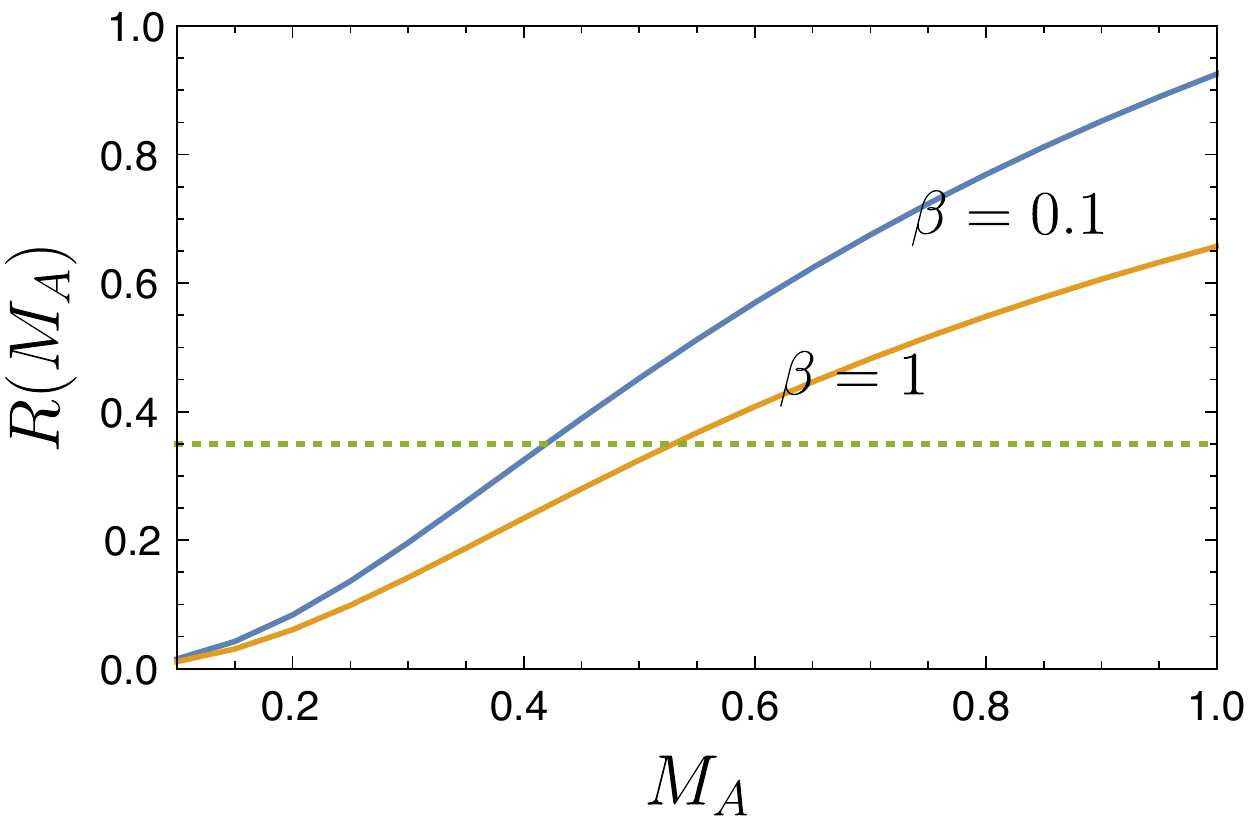}
\caption{Ratio of $B$ to $E$ power for synchrotron polarisation assuming an equal mix of Alfv\'en and slow modes for $\gamma_{\text{sync}}=0$. The dotted line shows the expected ratio of 0.35.}
\label{figmix}
\end{figure}

Looking at Fig. \ref{figsync}, one can see that the observed B/E ratio can be achieved by Alfv\'en MHD mode if the turbulence is sufficiently sub-Alfv\'enic with $M_A\lesssim 0.4$. This is consistent with the results from dust polarisation studies (\citetalias{kandel2017can}). Moreover, since slow modes always give a ratio less than the observed ratio, the constraint for Alfv\'en Mach number gets more relaxed if one considers a mix of Alfv\'en and slow modes, as shown in Fig. \ref{figmix}.

\begin{figure*}
\centering
\includegraphics[scale=0.4]{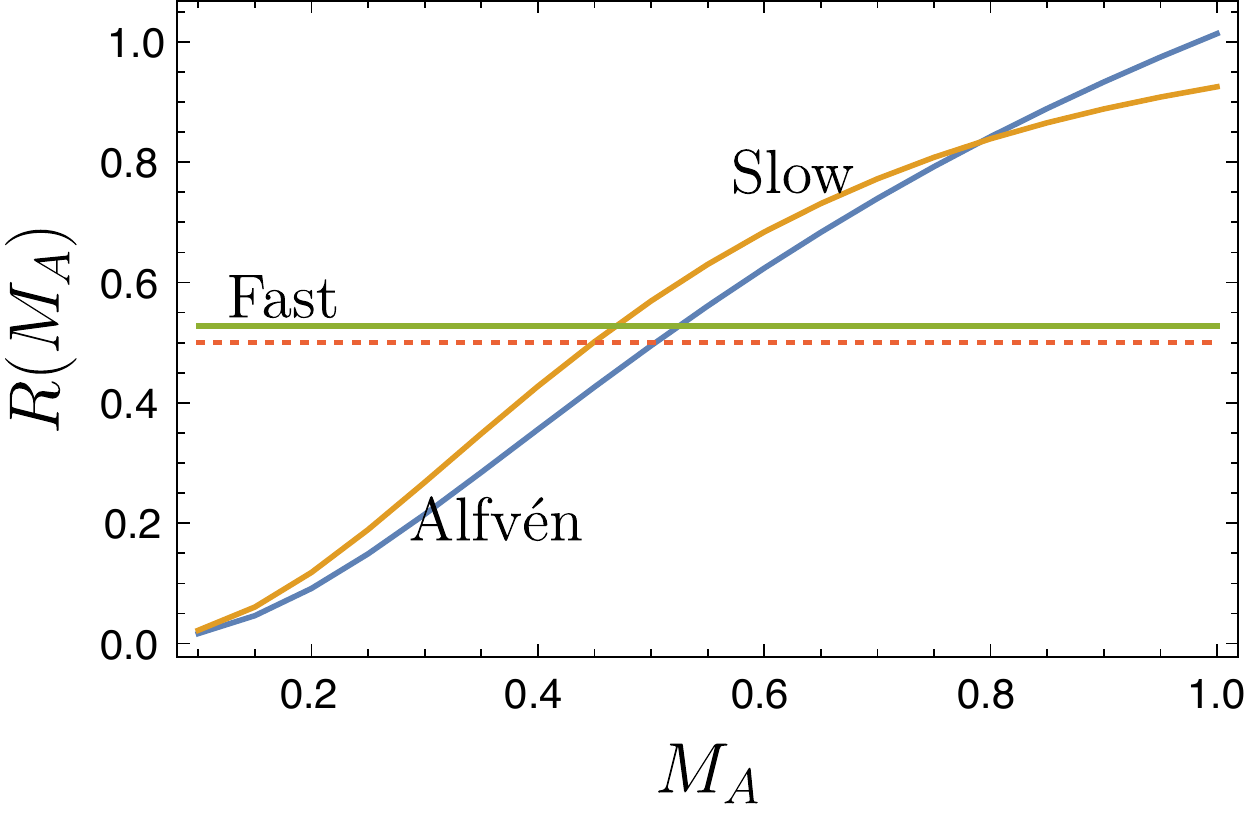}\hspace*{0.1cm}
\includegraphics[scale=0.4]{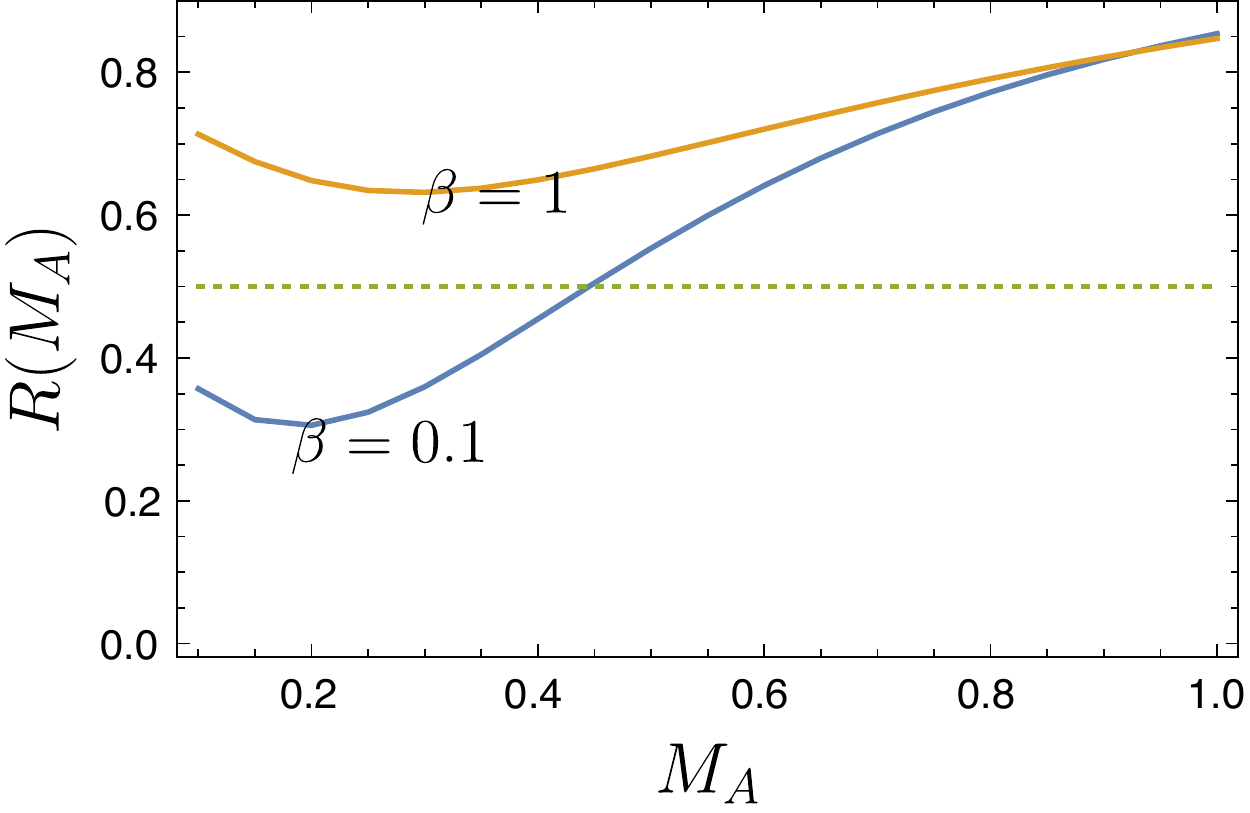}\hspace*{0.1cm}
\includegraphics[scale=0.4]{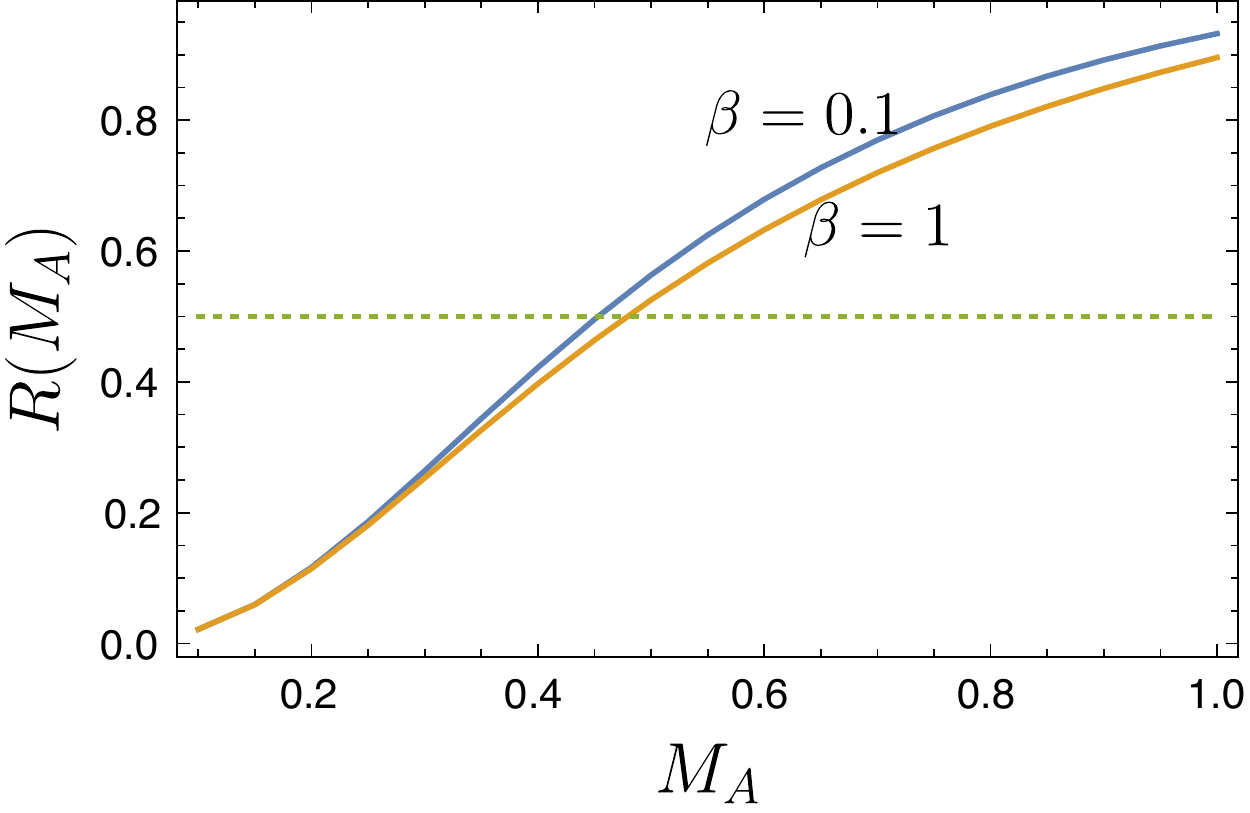}
\caption{Ratio of B/E power for dust polarisation for various $\beta$ at different Alfv\'en Mach number for $\gamma_d = -2$. Left: ratio due to each individual MHD modes. Center: ratio for an equal mix of power of three MHD modes. Right: ratio due to equal mix of Alfv\'en and slow only. The dotted line shows the observed ratio of 0.5 by {\it Planck}.}
\label{figebdust}
\end{figure*}

At higher frequencies, the dominant CMB foreground is dust. The ratio of E/B power for dust polarisation was presented in \citetalias{kandel2017can}. For completeness, here we repeat and present those results\footnote{Note that \citetalias{kandel2017can} of E/B power ratio. Here we calculate B/E power for consistency with synchrotron results.}. The amplitude of power of E and B polarisation mode and their ratio are calculated using Eqs. \eqref{main:gaesyncdust} - \eqref{fullequation}. Fig. \ref{figebdust} shows the B/E ratios for each MHD modes (left panel), for an equal mix of three MHD modes (central panel), and equal mix of Alfv\'en and slow modes (right panel). These plots suggest that the observed B/E ratio of 0.5 can be explained as long as turbulence is sub-Alfv\'enic, with $M_A\lesssim 0.5$, and as long as fast-modes are sub-dominant. These numbers are consistent with that inferred from synchrotron polarisation results. 

\subsection{TE cross-correlation}\label{sec:TE}
\begin{figure*}
\centering
\includegraphics[scale=0.5]{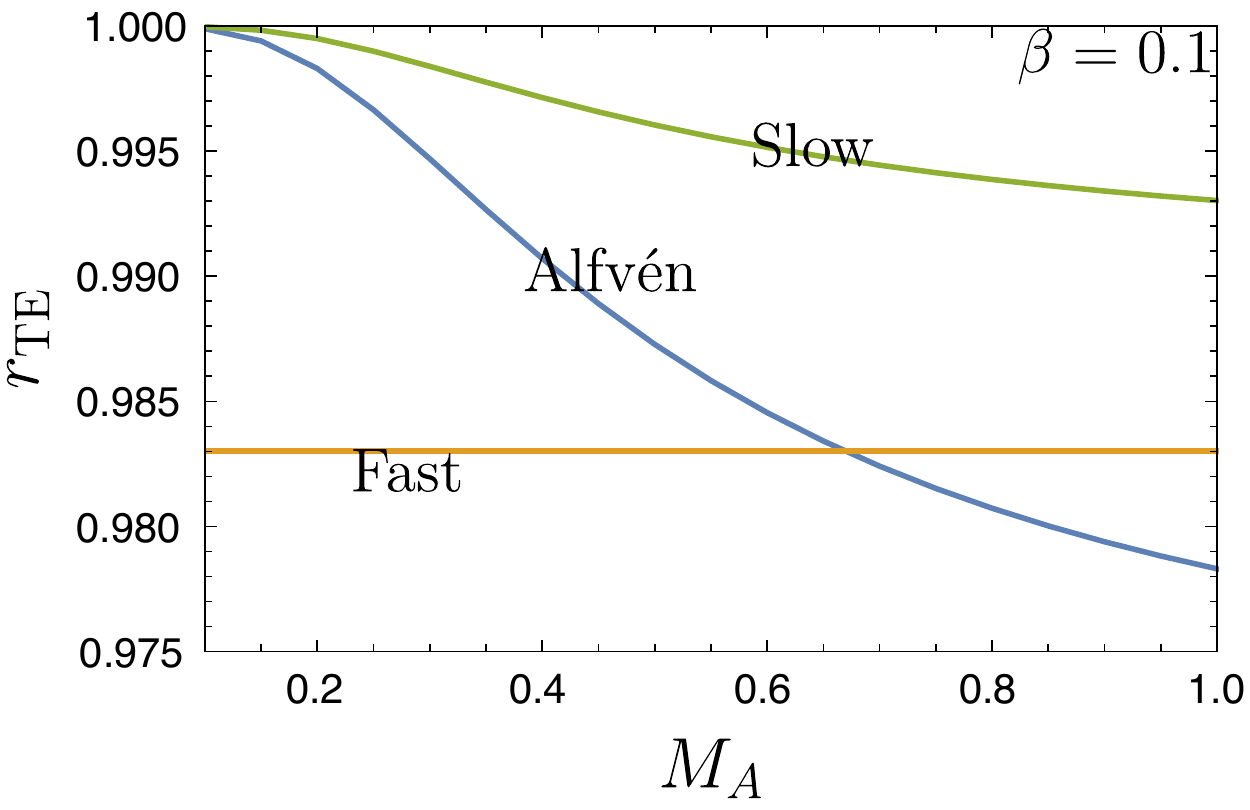}\hspace*{0.2cm}
\includegraphics[scale=0.5]{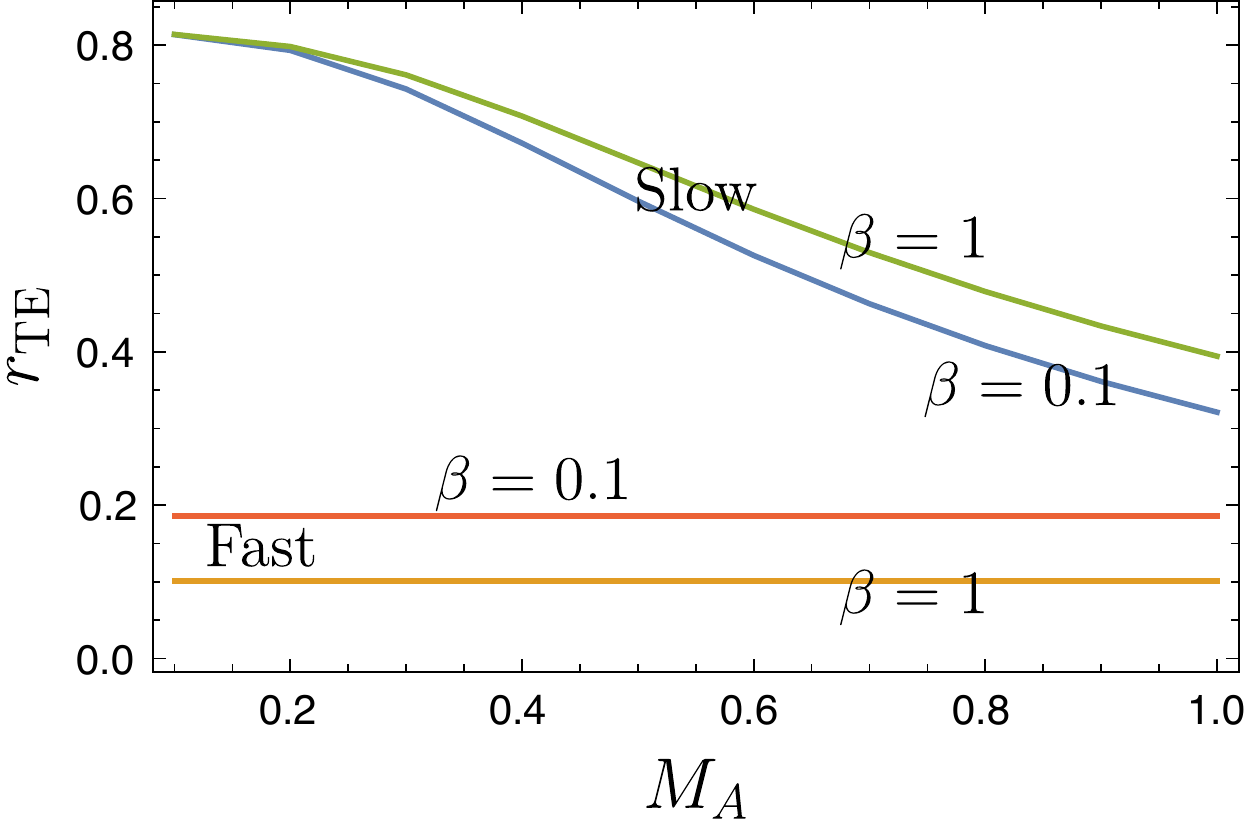}
\caption{Left: Synchrotron TE cross-correlation ratio for Alfv\'en, fast and slow modes for $\gamma_{\text{sync}}=1$ at $\beta=0.1$. All MHD modes contribute to TE cross-correlation, which is nearly unity for all three MHD modes. At the benchmark case $\gamma_{\text{sync}}=0$, TE cross-correlation for all MHD modes is 1. Right: Dust TE cross-correlation for fast and slow modes at $\gamma_{d}=-2$. Slow modes always give positive TE, whereas fast modes give a small negative values only at large $\beta$. Alfv\'en modes do not induce any temperature fluctuations, and thus do not have any TE cross-correlation. }
\label{figte}
\end{figure*}

We also compute TE cross-correlation for synchrotron polarisation using Eqs. \eqref{finaltesyncalfven} - \eqref{tecrossr}. First, Alfv\'en mode contributes to TE cross-correlation, which is distinct from dust polarisation, where Alfv\'en modes do not contribute to TE cross-correlation. Second, since both temperature fluctuations and polarisation fluctuations come from magnetic field fluctuations, we expect the TE cross-correlation to be rather high. In fact, if all looks at \eqref{finaltesyncalfven} - \eqref{cTTsync}, one can see that for $\gamma_{\text{sync}}=0$, the angular dependence of TE, TT and EE are identical for each MHD mode, resulting in a TE cross-correlation of 1. Moreover, the dependence of TE cross-correlation on $\gamma_{\text{sync}}$ is rather weak. This has been illustrated in the left-hand panel of Fig. \ref{figte}. As one can clearly see, the TE cross-correlation at $\beta=0.1$ and $\gamma_{\text{sync}}=1$ due to each MHD mode is $\sim 1$. Our numerical calculations also show that the TE cross-correlation for synchrotron is only weakly sensitive to plasma $\beta$. It would be interesting to compare this prediction with observational results, but unfortunately we are not aware of results on TE correlation of synchrotron foreground. 

Next, we compute the TE cross-correlation for dust polarisation using Eqs. \eqref{eq:dte} and \eqref{tecrossr}. Unlike in  synchrotron, Alfv\'en modes do not induce any temperature fluctuations and thus do not contribute to TE cross-correlation. The TE cross-correlation for fast and slow modes are shown in the right-hand panel of Fig. \ref{figte} at different plasma $\beta$. Our result shows that the TE  for slow modes is positive and $\gtrsim 0.4$ for all ranges of $\beta, M_A$, which is distinct from synchrotron result, which was negative for all range of parameters. Moreover, fast modes give a positive $TE$ for low $\beta$, while $TE\sim 0$ at $\beta\gg 1$. 
Our result is in very good agreement with {\it Planck} measurement of mean TE cross-correlation $\sim 0.36$ for all sky regions and for $5\lesssim\ell\lesssim 100$ (see Fig 4 of \citealt{akrami2018planck}). The only requirement for the consistency between our result and {\it Planck} measurement is that fast mode be sub-dominant, as fast mode give TE cross-correlation which is either lower than $\sim 0.36$ as measured by {\it Planck} or very close to zero. We stress that in our model, the same condition of sub-dominant fast mode is also required (see Sec. \ref{sec:EBpower}) to explain B/E power ratio observed by {\it Planck}. To sum, if one considers mix of modes, our result supports the observational results provided by {\it Planck} that TE cross-correlation is positive and about $\sim 0.36$. 

Looking at the right-hand panel of Fig. \ref{figte}, one can also infer several statistical properties of MHD turbulence. First, measuring a large positive TE cross-correlation (as indicated by \citetalias{caldwell2017dust}) implies dominance of slow and Alfv\'en modes, as well as suggest $M_A$ less than unity. Second, if observations show $TE\approx 0$, one can infer that the slow mode contribution to the turbulence is negligible, and only Alfv\'en and fast modes could be important modes. Thus, TE cross-correlation together with E/B power ratio, can be used to infer the dominance of each MHD modes, and constrain important parameters of the turbulence like $\beta, M_A$. 

Looking at dust TE cross-correlation, we found that unlike E/B power ratio, TE cross-correlation is sensitive to correlation between magnetic field and density. Assuming negligible correlation between magnetic field and density, we were able to match theory with {\it Planck} measurements. This is a  serious physical difference from \citetalias{caldwell2017dust}, which assumes the perfect flux-freezing, which had been for decades the paradigm for all astrophysical research. However, the flux-freezing assumption based on the \cite{alfven1942existence} theorem is not valid in turbulent fluids. Turbulent reconnection theory \citep{lazarian1999reconnection} predicts fast magnetic reconnection that is incompatible with the flux-freezing. This violation has been demonstrated numerically in \cite{eyink2013flux}. The Zeeman splitting observations in \cite{crutcher2010magnetic} show that up to certain threshold density there is no magnetic field-density correlation that was assumed in \citetalias{caldwell2017dust}. This threshold density is related to the development of the fast gravitational collapse of clumps \citep{lazarian2012magnetization} and is much higher than the densities that we encounter at high galactic latitudes. It is assuming no magnetic field-density correlation that we can account for the observed value of the TE correlations. In fact, the reported TE correlations can be used as yet another argument in favor of the flux-freezing violation in turbulent fluids (see more examples in a review \citealt{lazarian2015turbulent}). 

\subsection{Spectrum of power of E and B polarisation modes}
Although we focused on B/E power ratio as well as TE cross-correlation of polarisation signals, spectral index of $E$ and $B$ power is another important quantity of study. A simple analysis using Eq. \ref{maincell} shows that for Alfv\'en mode dominated turbulence, the spectrum of $E$ and $B$ powers for synchrotron polarisation should be close to Kolmogorov index, i.e. $P_E(\ell), P_B(\ell) \propto \ell^{-11/3}$. However, this result is valid only for large $\ell$, where the flat sky approximation is applicable. At small $\ell$, the spectrum is expected to be shallower. \citetalias{caldwell2017dust} noted that the spectrum of $E$ and $B$ mode power for dust polarisation observed by {\it Planck} is about $\sim 2.42$, different from the expected Kolmogorov index of $\sim 3.67$. {\it Planck} measurement of synchrotron polarisation also reveals spectral index approximately the same as for dust polarisation, which is again different from expected Kolmogorov index. However, as argued in \citet{cho2010galactic}, this disparity in the spectral index might be due to the non-trivial geometry of observations with the sampling being carried out along the diverging lines of sight, and within the volume where the density of emitters changes. Moreover, \citet{cho2010galactic} predicted that for the stratified model of Galactic dust, spectral index of $E$ and $B$ power approaches the Kolmogorov value of $-11/3$ for $\ell > 1000$, whereas for $\ell<1000$, the spectrum should be shallower. In fact, the E and B power {\it Planck} 353 GHz spectra computed down to $\ell=2$ indicates spectral flattening at $\ell<20$ (see Fig 20 of \citealt{aghanim2016planck}; also see \citealt{vansyngel2017statistical}), consistent with our argument. However, {\it Planck} spectra are dominated by noise variance at $\ell>1000$, and thus do not have full sensitivity to test whether the E and B power spectra follow Kolmogorov index at these scales. 

Another argument presented in \citet{cho2010galactic} is that if the synchrotron emissivity in the Galaxy depends on the distance from Galactic plane, or if the geometry of lines of sight is converging, as opposed to the usually adopted geometry of parallel lines of sight, the spectral index of $E$ and $B$ powers at low $\ell$ will be shallower than the Kolmogorov index. Therefore, the spectrum of polarisation powers observed by {\it Planck} for both dust and synchrotron are not inconsistent with the results of \citet{cho2010galactic}, as {\it Planck} probed scales of $\ell<1000$. 

\section{Discussion and summary}\label{discuss}

In this paper, we have extended our previous work from \citetalias{kandel2017can}, and shown how one can study MHD turbulence in the ISM using CMB foreground polarisation maps: both dust and synchrotron. The paper presents the B/E power ratio as well as TE cross-correlation as a function of $M_A$ and plasma $\beta$, for both synchrotron and dust polarisation, and compares the results to {\it Planck} findings. By taking a model with physical parameters $M_A$ and $\beta$, and by assuming negligible correlation between density and magnetic field, we showed that the B/E asymmetry and a positive TE cross-correlation both match well with {\it Planck} measurements. 

Our work was originally motivated by findings of \citetalias{caldwell2017dust} that the observed asymmetry in E and B power for dust polarisation as found by {\it Planck} is in tension with the existing theoretical model of MHD turbulence. As we have shown, this is not the case, and that the observed asymmetry in E and B power for not just the dust but also for the synchrotron polarisation is consistent with the existing model of MHD turbulence. However, we stress that the difference in our findings and that of \citetalias{caldwell2017dust} comes not from the difference in the analytical framework, but from the basic parameters in the problem that are used to interpret results. Here we discuss in more detail what the differences in the two paper are and how those affect the results for dust polarisation, and the interpretation of those results. First, our model for anisotropic power spectrum of the magnetic field is built upon physically observable parameter Alfv\'en Mach numnber $M_A$, unlike \citetalias{caldwell2017dust}, who use a $\lambda$ parameter for the characterisation of anisotropy. It is not immediately clear how $\lambda$ translates to physical observable, and thus it is not easy to interpret  whether a required value of $\lambda$ is realistic. Second, our model assumes uncorrelated density and the magnetic field, whereas \citetalias{caldwell2017dust} assume a full correlation between the two. While this assumption does not significantly affect the results of B/E power ratio, level of correlation between density and the magnetic field is important in determining TE cross-correlation. By comparing our results for dust TE cross-correlation with that of \citetalias{caldwell2017dust}, one can immediately see that while there are regions in parameter space $\lambda, \beta$ of \citetalias{caldwell2017dust} where TE cross-correlation is negative, our model gives a positive TE cross-correlation over all range of parameter space $M_A, \beta$. Clearly, our model matches better with the {\it Planck} prediction of positive TE cross-correlation. Moreover, as we have explained in Sec. \ref{sec:TE}, the numerical prediction of our model is very consistent with the {\it Planck} result of TE cross-correlation for dust.

Although we have argued that the asymmetry in E and B power and TE correlation for both dust and synchrotron polarisation arise due to MHD turbulence in the diffuse ISM, there are several works which have tried to address the B/E asymmetry as well as TE correlation for galactic foregrounds. The work in \cite{2016A&A...586A.141P} suggests that the dust B/E asymmetry and TE correlation arise mainly from the alignment of cold neutral medium (CNM) filamentary structures with the magnetic field. Although this is also a likely explanation, we would like stress that our model based on MHD turbulence is already sufficient to explain the observed {\it Planck} results. Similarly, there have been several numerical works that have tried to model the B/E asymmetry and spectral index of E and B power. For example, \cite{vansyngel2017statistical} have presented a method to simulate maps of polarised dust emission by modeling Galactic magnetic field as a superposition of a mean uniform field and a Gaussian random turbulent field with a power-law type power spectrum. Within their model, they were able to successfully explain B/E asymmetry as well as obtain the best fit for spectral indices of E and power power consistent with the {\it Planck} result. However, their model is just mathematical and cannot be used to constrain the physical origin of B/E asymmetry and TE correlation. 

The formalism we have developed in this paper will be useful in CMB foreground separation for cosmological studies. However, our formalism can also be applied to polarisation data from individual turbulent clouds to study their turbulence properties. This will be a valuable addition to the existing techniques which rely mostly on emission and absorption lines (see \citealt{kandel2016study}). Several degeneracies like angle between the magnetic field and the LOS, Alfv\'en Mach number, plasma $\beta$, and composition of individual MHD modes exist in these studies which use emission and absorption lines. As we have shown in this paper, both dust polarisation and synchrotron polarisation data are valuable tools to break these degeneracy. Future works will likely show the rich understanding of MHD turbulence one can extract using polarisation data.

The main results of the paper are listed below.

\begin{itemize}
    \item The {\it Planck} result of B/E power ratio for dust and synchrotron polarisation is not inconsistent with the  theoretical picture of MHD turbulence as long as turbulence in the ISM is sub-Alfv\'enic, with $M_A\lesssim 0.5$ and fast modes are sub-dominant. This is consistent with the expectation of sub-Alfv\'enic turbulence in high Galactic latitudes.
    \item The TE cross-correlation of $\sim 0.82$ for dust predicted by {\it Planck} fits with our model calculation, which assumes uncorrelated density-magnetic field. Thus, the reported TE correlations can be used as yet another argument in favour of the flux-freezing violation in turbulent fluids in the high Galactic latitudes.
    \item The TE cross-correlation together with measurements of B/E power ratio is a powerful method to constrain $M_A$ and dominance of individual MHD modes. For example, as we showed in this paper, for dust polarisation, the contribution of fast modes in TE correlation at high $\beta$ is negligible, and thus measuring a positive TE correlation would imply the dominance of slow and Alfv\'en modes, whereas TE cross-correlation close to 0 would imply dominance of fast and Alfv\'en modes. 
\end{itemize}
 
\section*{Acknowledgements}
AL acknowledges the support the NSF grant DMS 1622353 and AST 1715754.

\bibliographystyle{mnras}
\bibliography{polarizationdraft}
\appendix
\section{E and B mode description: general procedure}\label{app:general}
Here we present necessary theoretical framework to study dust and synchrotron polarisation due to MHD turbulence, which consists three modes: incompressible Alfv\'en mode, and compressible fast and slow modes. We provide expressions for power of $E$ and $B$ polarisation modes due to each MHD cascade, in a general form that will be applicable to both dust and synchrotron polarisation with appropriate modifications.

\subsection{Synchrotron polarisation}
For synchrotron radiation the two-dimensional projection of an emitting volume has polarisation $\epsilon_{P, \text{sync}}=\epsilon_Q+i\epsilon_U$ and intensity $\epsilon_{I, \text{sync}}$ emissivities of the following form
\begin{eqnarray}\label{eq:intengen}
\epsilon_P&=&A_{\text{sync}, P} \;H_{\perp}^{\gamma_{\text{sync}}}(H_x+\mathrm{i}H_y)^2~,\\
\epsilon_I&=&A_{\text{sync}} \;
H_{\perp}^{\gamma_{ \text{sync}}}\lvert H_x+\mathrm{i}H_y\rvert^2\nonumber\\
&=&A_{I, \text{sync}}H_{\perp}^{\gamma_{\text{sync}}+2}~.
\end{eqnarray}

Here  $X = (x, y)$ is the 2D position vector on the sky, 
$A_{\text{sync}, P}$ and $A_{\text{sync}, I}$ are amplitudes that depend on several parameters like number density of relativistic electrons and $H_\perp = \sqrt{H_x^2+H_y^2}$. The power $\gamma_{\text{sync}}$ of the
dependence on magnetic field strength, can be, in principle fractional and
depends on the spectrum of relativistic electrons. However $\gamma_{\text{sync}}=0$
describes the correlation properties of the system reasonably well (see \citealt{lazarian2012statistical}).

We now consider the fluctuation of polarised emission. In the coordinate
frame where $x$-axis is aligned with sky projection of the mean
magnetic field, $\mathbf{H_0}=H_0(\sin\theta, 0, \cos\theta)$,
\begin{align}
\delta\epsilon_P=A_PH_0^{\gamma_{\text{sync}}+2}(\sin\theta)^{\gamma_{\text{sync}}+1}\left(2\frac{\delta H_x+i\delta H_y}{H_0}+\gamma_{\text{sync}}\frac{\delta H_\perp}{H_0}\right)~,
\end{align}
where $\delta H_\perp=\delta H_x$. Using Eq. \eqref{eq:transform}, it can be shown that the E and B amplitude in a given Fourier mode of wavevector $\bm{k}$ transverse to the LOS in a box of radial width $\Delta r$ have a form 
\begin{align}
\tilde{E}+i\tilde{B}=An_0H_0^{\gamma_{\text{sync}}+2}\frac{\Delta r}{r^2}\mathrm{e}^{-2i\psi}(\sin\theta)^{\gamma_{\text{sync}}+1}\bigg(2\frac{\delta \tilde{H}_x+i\delta \tilde{H}_y}{H_0}\nonumber\\
+\gamma_{\text{sync}}\frac{\delta \tilde{H}_\perp}{H_0}\bigg)~,
\end{align}
where $\tilde{H}_x$ is the Fourier transform of the x-component of magnetic field  and so on. With the above equation, we finally have
\begin{align}
\tilde{E}=An_0H_0^{\gamma_{\text{sync}}+2}\frac{\Delta r}{r^2}(\sin\theta)^{\gamma_{\text{sync}}+1}\bigg((2+\gamma_{\text{sync}})\cos 2\psi\frac{\delta\tilde{H}_x}{H_0}\nonumber\\+2\sin 2\psi\frac{\delta\tilde{H}_y}{H_0}\bigg)~,
\end{align}
and 
\begin{align}
\tilde{B}=An_0H_0^{\gamma_{\text{sync}}+2}\frac{\Delta r}{r^2}(\sin\theta)^{\gamma_{\text{sync}}+1}\bigg(-(2+\gamma_{\text{sync}})\sin 2\psi\frac{\delta\tilde{H}_x}{H_0}\nonumber\\+2\cos 2\psi\frac{\delta\tilde{H_y}}{H_0}
\bigg)~,
\end{align}
Re-writing the magnetic-field perturbations in-terms of two transverse-vector modes, those in Alfv\'en wave and in slow and fast waves direction, we have 
\begin{align}\label{esync}
\tilde{E}=-A_{\text{sync}}H_0^{\gamma_{\text{sync}}+2}\frac{\Delta r}{r^2}(\sin\theta)^{\gamma_{\text{sync}}}\frac{\left(2-\gamma_\text{sync}\cos 2 \psi\right)}{\sin\alpha}\nonumber\\
\times\sin\theta\sin\psi\bigg(\cos\theta
\frac{\delta\tilde{H}_a}{H_0}
+\sin\theta \sin\psi  \frac{\delta\tilde{H}_p}{H_0}\bigg)~,
\end{align}
and
\begin{align}\label{newbsync}
\tilde{B}=-A_{\text{sync}}H_0^{\gamma_{\text{sync}}+2}\frac{\Delta r}{r^2}(\sin\theta)^{\gamma_{\text{sync}}}\frac{\left(2+2\gamma_\text{sync}\sin^2\psi\right)}{\sin\alpha}\nonumber\\
\times\sin\theta\cos\psi\bigg(\cos\theta 
\frac{\delta\tilde{H}_a}{H_0}+\sin\theta\sin\psi\frac{\delta\tilde{H}_p}{H_0}\bigg)~,
\end{align}
where subscript $a$ represents Alfv\'en mode and $p$ represents compressible modes.

Using Eqs. \eqref{esync} and \eqref{newbsync}, we finally have for Alfv\'en mode
\begin{align}\label{eq:gaesyncm}
\langle\tilde{E}^2\rangle_{\text{sync},a}
\propto(\sin\theta&)^{2\gamma_{\text{sync}}+2}\cos^2\theta\sin^2\psi\nonumber\\
&\times\left(\frac{2-\gamma_{\text{sync}}\cos2\psi}{\sin\alpha}\right)^2P_{a,H}(\alpha)~,
\end{align}
and
\begin{align}\label{eq:gabsync}
\langle\tilde{B}^2\rangle_{\text{sync},a}\propto(\sin\theta&)^{2\gamma_{\text{sync}}+2}\cos^2\theta\cos^2\psi\nonumber\\
&\times\left(\frac{2+2\gamma_{\text{sync}}\sin^2\psi}{\sin\alpha}\right)^2P_{a,H}(\alpha)~.
\end{align}
Similarly, for slow and fast modes, we have
\begin{align}\label{gsfm}
\langle\tilde E^2\rangle_{\text{sync},i}\propto(\sin\theta&)^{2\gamma_{\text{sync}}+4}\sin^4\psi\nonumber\\
&\times\left(\frac{2-\gamma_{\text{sync}}\cos2\psi}{\sin\alpha}\right)^2P_{i,H}(\alpha)~,
\end{align}
and
\begin{align}\label{gsfmb}
\langle\tilde B^2\rangle_{\text{sync},i}\propto(\sin\theta&)^{2\gamma_{\text{sync}}+4}\sin^2\psi\cos^2\psi\nonumber\\
&\times\left(\frac{2+2\gamma_{\text{sync}}\sin^2\psi}{\sin\alpha}\right)^2P_{i,H}(\alpha)~.
\end{align}

\subsection{Dust}
The formalism for dust emission has been presented in \citetalias{caldwell2017dust}. Here we reproduce the necessary steps both for completeness of exposition and for comparison with the case of synchrotron. The polarised emmisivity and intensity of the dust is taken
in the form
\begin{eqnarray}\label{eq:intengendust}
\epsilon_P&=&A_{\text{d},P} \;nH^{\gamma_{\text{d}}}(H_x+\mathrm{i}H_y)^2~,\\
\epsilon_I&=&A_{\text{d},I}\;n~.
\end{eqnarray}
with the degree of polarisation
\begin{equation}
p \equiv \frac{|\epsilon_P|}{\epsilon_I} = \frac{A_{ \text{d}, P}}{A_{\text{d},I}}H^{\gamma_{\text{d}+2}}~.
\end{equation}
Here $n$ is the dust particles number density, 
$A_{\text{d}, P}$ and $A_{\text{d}, I}$ are amplitudes of the emissivity. and  $H = \sqrt{H_x^2+H_y^2+H_z^2}$ is the magnetic field magnitude. 
The power index $\gamma_{\mathrm{d}}$ describes dependence of the 
polarisation level on the strength of the magnetic field. The benchmark
case is $\gamma_{\mathrm{d}}=-2$ when this dependence is absent and
the magnetic field defines only the direction of the polarisation.

Our descriptions of dust and synchrotron emission have many formal similarities, but let us point out the major differences. Firstly, the intensity of synchrotron is directly determined by the
strength of the magnetic field, while intensity of dust emission
is primarily determined by the dust particle density.
Secondly, the density
of the dust emitters fluctuates in the turbulent medium, while the 
density of synchrotron emitting relativistic  electrons is taken constant.
For dust, density fluctuations contribute additional source of 
fluctuations both to polarisation and intensity measures and their correlations. Lastly, we expect different values for the power indexes
$\gamma_\text{sync}$ and $\gamma_\text{d}$.

Let us now consider the fluctuations in the polarised emissivity:
\begin{align}
\delta\epsilon_d=A_d n_0H_0^{\gamma_{\text{d}}+2}\bigg(2\sin\theta\frac{\delta H_x+i\delta H_y}{H_0}+\gamma_{\text{d}}\sin^2\theta\frac{\delta H}{H_0}\nonumber\\
+\sin^2\theta\frac{\delta n}{n}\bigg)~,
\end{align}
where $\delta H=\sin\theta\,\delta H_x+\cos\theta\,\delta H_z$. Using Eq. \eqref{eq:transform}, it can be shown that the E and B amplitude in a given Fourier mode of wavevector $\bm{k}$ transverse to the LOS in a box of radial width $\Delta r$ have a form (see \citetalias{caldwell2017dust})
\begin{align}
\tilde{E}+i\tilde{B}=A_dn_0H_0^{\gamma_{\text{d}}+2}\frac{\Delta r}{r^2}\mathrm{e}^{-2i\psi}\bigg(2\sin\theta\frac{\delta \tilde{H}_x+i\delta \tilde{H}_y}{H_0}\nonumber\\
+\gamma_{\text{d}}\sin^2\theta\frac{\delta \tilde{H}}{H_0}+\sin^2\theta\frac{\delta\tilde{n}}{n_0}\bigg)~,
\end{align}
where $\tilde{H}_x$ is the Fourier transform of the x-component of magnetic field  and so on. With the above equation, we finally have
\begin{align}
\tilde{E}=A_dn_0H_0^{\gamma_{\text{d}}+2}\frac{\Delta r}{r^2}\bigg(2\sin\theta\cos 2\psi\frac{\delta\tilde{H}_x}{H_0}+2\sin\theta\sin 2\psi\frac{\delta\tilde{H}_y}{H_0}\nonumber\\
+\gamma_{\text{d}}\sin^2\theta\cos 2\psi\frac{\delta\tilde{H}}{H_0}+\sin^2\theta\cos 2\psi\frac{\delta\tilde{n}}{n_0}\bigg)~,
\end{align}
and 
\begin{align}
\tilde{B}=A_dn_0H_0^{\gamma_{\text{d}}+2}\frac{\Delta r}{r^2}\bigg(-2\sin\theta\sin 2\psi\frac{\delta\tilde{H}_x}{H_0}+2\sin\theta\cos 2\psi\frac{\delta\tilde{H_y}}{H_0}\nonumber\\
-\gamma_{\text{d}}\sin^2\theta\sin 2\psi\frac{\delta\tilde{H}}{H_0}-\sin^2\theta\sin 2\psi\frac{\delta\tilde{n}}{n_0}\bigg)~,
\end{align}
Re-writing the magnetic-field perturbations in-terms of two transverse-vector modes, those in Alfv\'en wave and in slow and fast waves direction, we have (see \citetalias{caldwell2017dust})
\begin{align}\label{esyncdust}
&\tilde{E}=A_dn_0H_0^{\gamma_{\text{d}}+2}\frac{\Delta r}{r^2}\bigg(-\frac{\delta\tilde{H}_a}{H_0}\sin 2\theta\frac{\sin\psi}{\sin\alpha}\nonumber\\
&-\frac{\sin^2\theta(1-\cos2\psi(1+\gamma_{\text{d}}\sin^2\alpha))}{\sin\alpha}\frac{\delta\tilde{H}_p}{H_0}\nonumber\\
&+\sin^2\theta\cos 2\psi\frac{\delta\tilde{n}}{n_0}\bigg)~,
\end{align}
and
\begin{align}
\tilde{B}&=A_dn_0H_0^{\gamma_{\text{d}}+2}\frac{\Delta r}{r^2}\bigg(-\frac{\delta\tilde{H}_a}{H_0}\sin 2\theta\frac{\cos\psi}{\sin\alpha}\nonumber\\
&-\frac{\sin^2\theta\sin2\psi\left(1+\gamma_{\text{d}}\sin^2\alpha\right)}{\sin\alpha}\frac{\delta\tilde{H}_p}{H_0}-\sin^2\theta\sin 2\psi\frac{\delta\tilde{n}}{n_0}\bigg)~,
\end{align}
where subscript $a$ represents Alfv\'en mode and $p$ represents compressible modes.

Alfv\'en modes do not induce any density fluctuations and therefore they give
\begin{equation}\label{eq:gaesyncdust}
\langle\tilde{E}^2\rangle_{d,a}\propto\sin^2 2\theta\frac{\sin^2\psi}{\sin^2\alpha}P_{a,H}(\alpha)~,
\end{equation}
and
\begin{equation}\label{eq:gabsyncdust}
\langle\tilde{B}^2\rangle_{d,a}\propto\sin^2 2\theta\frac{\cos^2\psi}{\sin^2\alpha}P_{a,H}(\alpha)~.
\end{equation}

For slow and fast modes, assuming uncorrelated density and magnetic field, we have
\begin{align}\label{gsfmdust}
\langle\tilde E^2\rangle_{d,i}\propto  \frac{\sin^4\theta(1-\cos2\psi(1+\gamma_{\text{d}}\sin^2\alpha))^2}{\sin^2\alpha}P_{i, H}(\alpha)\nonumber\\
+\sin^4\theta\cos^22\psi P_{i, \rho}(\alpha)~,
\end{align}
and
\begin{align}\label{gsfmbdust}
\langle\tilde B^2\rangle_{d,i}\propto \frac{\sin^4\theta\sin^22\psi(1+\gamma_{\text{d}}\sin^2\alpha)^2}{\sin^2\alpha}P_{i, H}(\alpha)\nonumber\\
+\sin^4\theta\sin^22\psi P_{i, \rho}(\alpha)~.
\end{align}

\subsection{Quantities of Interest}

The main quantity of interest for us is  of power in E mode, which we define to be 
\begin{equation}\label{eq:powerE}
P_{i,E}=\int\mathrm{d}\Omega\,\tilde{E}^2~,
\end{equation}
and power in B mode defined as
\begin{equation}\label{eq:powerB}
P_{i,B}=\int\mathrm{d}\Omega\,\tilde{B}^2~,
\end{equation}
and the ratio of B to E power
\begin{equation}\label{eq:powerBEm}
R=\frac{\int\mathrm{d}\Omega\tilde{B}^2}{\int\mathrm{d}\Omega\tilde{E}^2}~.
\end{equation}
This ratio has been obtained from the {\it Planck} survey at high galactic latitudes for both dust and synchrotron. In Eqs. \eqref{eq:powerE}-\eqref{eq:powerBEm}, each integral represent three dimensional angular averaging. 

Eqs. \eqref{eq:powerE}-\eqref{eq:powerBEm} are only applicable for individual MHD modes. In reality, MHD turbulence contains mixture of MHD modes, each with different wave-spectrum, and therefore in such cases, one has to use the full expression (c.f Eq. \eqref{maincell})
\begin{equation}\label{fullequation2}
R=\frac{\int\mathrm{d}r\,\mathrm{d}\Omega\,\tilde{B}^2(k=\ell/r)/r^2}{\int\mathrm{d}r\,\mathrm{d}\Omega\,\tilde{E}^2(k=\ell/r)/r^2}~.
\end{equation}
Assuming equal power in all three MHD modes at injection scale, one can obtain a following expression for B/E power ratio
\begin{equation}\label{fullequation1}
R=\frac{P_{a,B}+P_{s,B}+P_{f,B}\ell^{-1/6}}{P_{a,E}+P_{s,E}+P_{f,E}\ell^{-1/6}}~,
\end{equation}
where $P_{a,E}, P_{a,B}$ are the amplitudes of power of $E$ mode and $B$ mode, respectively, of Alfv\'en mode one obtains after carrying out angular integration of power spectrum, and so on for fast and slow modes. Eq. \ref{fullequation1} shows that the E/B power ratio is weakly sensitive to $\ell$, and for most cases one can effectively ignore the $\ell$ dependence all together. 

To simplify evaluation of all the angular integration, we transform $\{\theta, \psi\}\rightarrow\{\varpi, \alpha\}$ through $\cos\theta=\sin\alpha\cos\varpi, \sin\theta\sin\psi=\sin\alpha\sin\varpi$, and $\sin\theta\cos\psi=\cos\alpha$, such that integration is carried over $\mathrm{d}\Omega=\sin\alpha\,\mathrm{d}\alpha\,\mathrm{d}\varpi$.

\bsp	
\label{lastpage}

\end{document}